\documentclass[12pt,preprint]{aastex}

\begin{document}

\shorttitle{X-ray flares in Orion stars. I.} \shortauthors{Getman
et al.} \slugcomment{Accepted for publication in the Astrophysical Journal 07/11/08}

\title{X-ray flares in Orion young stars. I. Flare characteristics}

\author{Konstantin V.\ Getman\altaffilmark{1}, Eric D.\
Feigelson\altaffilmark{1}, Patrick S.\ Broos\altaffilmark{1},
Giuseppina Micela\altaffilmark{2}, Gordon P.\
Garmire\altaffilmark{1}}

\altaffiltext{1}{Department of Astronomy \& Astrophysics, 525
Davey Laboratory, Pennsylvania State University, University Park
PA 16802} \altaffiltext{2}{INAF, Osservatorio Astronomico di
Palermo G. S. Vaiana, Piazza del Parlamento 1, I-90134 Palermo,
Italy}

\email{gkosta@astro.psu.edu}

\begin{abstract}
Pre-main sequence (PMS) stars are known to produce powerful X-ray
flares which resemble magnetic reconnection solar flares scaled by
factors up to $10^4$. However, numerous puzzles are present
including the structure of X-ray emitting coronae and
magnetospheres, effects of protoplanetary disks, and effects of
stellar rotation. To investigate these issues in detail, we
examine 216 of the brightest flares from 161 PMS stars observed in
the Chandra Orion Ultradeep Project (COUP). These constitute the
largest homogeneous dataset of PMS, or indeed stellar flares at
any stellar age, ever acquired. Our effort is based on a new flare
spectral analysis technique that avoids nonlinear parametric
modeling. It can be applied to much weaker flares and is more
sensitive than standard methods. We provide a catalog with $>30$
derived flare properties and an electronic atlas for this unique
collection of stellar X-ray flares. The current study (Paper I)
examines the flare morphologies, and provides general comparison
of COUP flare characteristics with those of other active X-ray
stars and the Sun. Paper II will concentrate on relationships
between flare behavior, protoplanetary disks,  and other stellar
properties.

Several results are obtained. First, the COUP flares studied here
are among the most powerful, longest, and hottest stellar X-ray
flares ever studied. Peak luminosities are in the range $31< \log
L_{X,pk}< 33$ erg~s$^{-1}$; rise (decay) timescales range from
1~hour to 1~day (few hours to 1.5 days); many peak temperatures
exceed 100~MK. The scale of their inferred associated coronal
structures is $0.5-10$~$R_{\star}$. Second, no significant
statistical differences in peak flare luminosity or temperature
distributions are found among different morphological flare
classes, suggesting a common underlying mechanism for all flares.
Third, comparison with the general solar-scaling laws indicates
that COUP flares may not fit adequately proposed power-temperature
and duration-temperature solar-stellar fits. Fourth, COUP
super-hot flares are found to be brighter but shorter than cooler
COUP flares. Fifth, the majority of bright COUP flares can be
viewed as enhanced analogs of the rare solar ``long-duration
events''.
\end{abstract}

\keywords{open clusters and associations: individual (Orion Nebula
Cluster) - stars: flare - stars: pre-main sequence - X-rays:
stars}

\section{INTRODUCTION \label{introduction_section}}

All solar-type stars exhibit their highest levels of magnetic
activity during their pre-main sequence (PMS) phase
\citep{Feigelson07}. This includes `superflares' with peak
luminosities $\log L_x \ga 32$ erg s$^{-1}$ in the $0.5-8$ keV
band, $10^4$ more powerful than the strongest flares seen in the
contemporary Sun \citep[e.g.][]{Tsuboi98, Grosso04, Favata05}. PMS
stars thus join RS~CVn binary systems \citep[e.g.][]{Osten07} as
laboratories to study the physics of the most powerful magnetic
reconnection events.  PMS stars are more distant and fainter than
the closer RS~CVn systems, but hundreds of flaring PMS stars can
be simultaneously studied due to their concentration in rich
clusters.

The magnetic field structure of PMS stars, and thus the nature of
their reconnection and flaring, may (or may not) qualitatively
differ from other stars due to the presence of a protoplanetary
disk during the early PMS stages. The intense high energy
radiation from these PMS reconnection events may affect the
physical and chemical properties of the surrounding circumstellar
environment and play an important role in the formation of planets
\citep{Glassgold05,Feigelson07}.  A consensus has emerged during
the past decade that PMS accretion is funneled by magnetic field
lines linking the disk inner edge to the stellar surface
\citep[e.g.][]{Hartmann98, Shu00}.  However, while early theory
assumed a dipolar field morphology, recent studies point to a
complex multipolar field structure similar to the Sun's
\citep{Jardine06, Donati07, Long08}.

It is also unclear whether the X-ray flares occur primarily in
large loops with both footprints anchored on the stellar surface,
or in loops linking the stellar photosphere with the inner rim of
the circumstellar disk \citep{Isobe03, Favata05}. The first case
may suffer instability due to centrifugal force \citep{Jardine99}
while the second case may load the loop with cool accreting
material so that X-rays may not be produced \citep{Preibisch05}.

The 13-day nearly continuous observation of $\sim 1408$ PMS stars
in the Orion Nebula, the Chandra Orion Ultradeep Project
\citep[COUP;][]{Getman05}, enables both studies of individual
flare properties and statistical studies of flaring from Orion
stars \citep{Wolk05, Flaccomio05, Stassun06, Caramazza07,
Colombo07}. COUP also provided a unique opportunity to study
relatively rare superflares and long-duration flares.
\citet{Favata05} have analyzed the strongest 32 flares in the COUP
dataset using a long-standing method of time resolved spectroscopy
(TRS) modeled as cooling plasma loops. They concluded that at
least 1/3 of these are produced by magnetic reconnection in very
long coronal $5-20$~R$_{\star}$ structures. Such structures were
predicted in magnetospheric accretion models \citep[e.g][]{Shu97}
but not clearly identified before COUP. \citet{Favata05}
recognized that their sample was too small to quantitatively probe
the relationship between long coronal flaring structure and disks
or accretion.

The aim of the current study is to extend the flare sample of
\citet{Favata05} utilizing a more sensitive technique of flare
analysis, the ``method of adaptively smoothed median energy''
(MASME) introduced by \citet{Getman06}.  We combine this method
with the astrophysical cooling loop models of \citet{Reale97} to
trace the evolution of the flare plasma in temperature-density
diagrams and derive flaring loop sizes. The method allowed us to
examine $216$ of the brightest flares from $161$ brightest COUP
PMS stars. These constitute the largest homogeneous sample of
powerful stellar flares ever acquired in the X-ray band. In 
\citet{Getman08} (Paper~II), we use these results to study in detail the relationships
between PMS X-ray flares, stellar properties, protoplanetary
disks, and accretion.

Our flare analysis and the derived flare properties and
classifications are presented in \S \ref{analysis}. Properties of
the stars themselves are also provided. Global properties of our
flares are considered in \S \ref{results} and compared to
published studies of older stars.

\section{FLARE ANALYSIS \label{analysis}}

We analyze 216 of the brightest X-ray flares from 161 brightest
COUP young stars. This is a $>5$-fold increase from the bright
flare sample of \citet{Favata05}. The major steps of our analysis
are presented below.

\subsection{Flare Sample \label{source_sample_analysis}}

Among 1616 COUP X-ray sources, 1408 have been associated with the
stellar members of the Orion region \citep{Getman05b}. With the
exception of 10 hot OB-type stars, 1398 are identified as cool
young members of the Orion region \citep{Feigelson05}.

We start with the sub-sample consisting of the brightest 188 cool
Orion stars, those with net counts $NC \geq 4000$ as tabulated by
\citet[][Table 4]{Getman05}. All of them show signs of
variability, i.e. their X-ray lightcurves are characterized by at
least two Bayesian Block segments (Table 7 of Getman et al.)
during the 13.2-day COUP observation. The complete set of 1616
COUP lightcurves are available in the Source Atlas provided as a
figure set in the electronic edition of Getman et al. Individual
examples of strong flares are published in \citet{Favata05} and
\citet{Wolk05}.

Using the interactive software graphical program $function\_1d$ in
the IDL-based TARA package\footnote{Description and code for the
``Tools for ACIS Review \& Analysis'' (TARA) package developed at
Penn State can be found at
\url{http://www.astro.psu.edu/xray/docs/TARA/}.}, we identify
flare-like events in which the peak count rate is $\geq 4$~times
that of the ``characteristic'' level. The characteristic level is
the typical pre-flare or inter-flare; as explained by
\citet{Wolk05}. We avoid the designation ``quiescent'' level
because it probably arises from the integrated effect of many
weaker flares. This results in 216 flares from 161 cool stars with
median of $\sim 1500$ counts. Only 20 flares have $<600$ counts.

\subsection{MASME Spectral Modeling  \label{sliding_kernel_method_analysis}}

Our analysis of flare spectral evolution is based on the method of
adaptively smoothed median energy (MASME) introduced by
\citet{Getman06}.  We compare its results to the long-standing
method of time-resolved spectroscopy (TRS), which involves
statistical fitting of multiparameter spectral models, in
Appendix~ \ref{app:TRS_analysis}. The MASME method is simpler,
statistically more stable, and computationally quicker than TRS.
It employs an adaptively smoothed estimator of the median energy\footnote{
X-ray source median energy is an observed quantity \citep{Getman05}  and can
be effectively used as an indicator of plasma temperature if absorbing column is
known, or indicator of column density at median energies $>1.7$~keV \citep{Feigelson05}.
Other researches have also found that median energies are 
effective spectral estimators in X-ray CCD spectroscopy \citep{Hong04}.}
of flare counts and count rate to infer the evolution of plasma
temperatures and emission measures during the flare. This
procedure is similar to the analyses of \citet{Reale98,Wargelin07}
which avoid nonlinear parametric flare modeling using hardness
ratios.

The smoothing kernel is a rectangle (``boxcar'') of variable
width; as it moves through the time series its width is adjusted
so that it encompasses a specified number of X-ray counts. Kernels
are evaluated on a time grid with bin width $\Delta t = 1.14$~ks,
chosen to divide the total COUP observation span of 1140~ks into
1000 bins. Flux and median energy estimates are computed from the
counts found in each of these overlapping kernels, forming
smoothed flux and median energy time series which then used for
spectral modeling. Similar flux and median energy time series are
produced by the
\anchor{http://www.astro.psu.edu/xray/docs/TARA/ae_users_guide.html
}{ACIS Extract} software package\footnote{ACIS Extract is a {\it
Chandra}-ACIS source extraction and analysis package developed at
Penn State. It was used for the COUP data analysis and is
described at
\url{http://www.astro.psu.edu/xray/docs/TARA/ae\_users\_guide.html}.}.
For each flare, the target number of counts in the kernel is
chosen such that the resulting smoothed light curve closely
matches the binned lightcurve given in the COUP Source Atlas of
Getman et al. (2005a). This generally results in $\sim 100$ to
$\sim 600$ counts included in each kernel for the weakest to the
brightest flares, respectively. The typical width of the smoothing
kernel is a few kiloseconds at the peak of the flare and $\sim
10$~ks at the base of the flare.

PMS X-ray spectra are typically modeled as emission by hot plasmas
at one or two temperatures subject to absorption by a column
density, $N_H$. If we adopt the $N_H$ inferred from the
time-integrated spectral fits of \citet{Getman05} and assume that
$N_H$ does not change during the COUP observation, then we can
calibrate median energies and count rates to plasma temperatures
($kT$) and emission measures ($EM$). Figure \ref{fig_kt_vs_mede}
shows the median energy to plasma temperature calibration based on
simulations similar to those described by \citet{Getman06}. For
each COUP source, the {\it fakeit} command in XSPEC
\citep{Arnaud96} is used to simulate a grid of high
signal-to-noise ACIS-I spectra with absorbed plasma models
(WABS$\times$MEKAL) and fixed column density. Simulated spectra
are then passed through ACIS Extract to perform photometric
analysis, including calculation of fluxes (observed count rates)
and median energy. At each time point of interest comparison of
simulated count rates with that of the observed flare provides an
estimate of flare emission measure; comparison of simulated and
observed median energies provides an estimate of flare plasma
temperature\footnote{XSPEC fitting of simulated spectra using the
$\chi^2$ statistic tends to overestimate plasma temperature. In
order to avoid this systematic bias in our simulations, we adopt
the input plasma temperature rather than the temperature derived
from XSPEC fit of the simulated spectrum.}. Due to the curved
dependency of the true plasma temperature on observed median
energy (Figure \ref{fig_kt_vs_mede}), estimated plasma
temperatures of $\la 200$~MK ($\la 17$~keV) are expected to be
quite reliable, while temperature values of $\ga 200$~MK are much
more uncertain. The accuracy of our temperature estimates is
discussed in detail in Appendix~\ref{app:MASME_errors}. We find
the 1$\sigma$ errors of peak flare temperature are generally
$<30\%$ for $T_{obs,pk} < 50$~MK, around $45\%$ for $T_{obs,pk}
\sim 200$~MK, but reach up to $65\%$ for the hottest flares.
Nonetheless, we establish that `superhot' flares with derived peak
temperatures above 200~MK are definitely hotter than 100~MK.

The major advantage of the MASME over TRS is that spectral
analysis is performed at dozens of points along the decay phase of
a flare light curve, rather than at just a handful of points
typical of TRS. This higher time resolution often results in the
discovery of a more detailed, often more complex flare behavior. A
drawback of the MASME procedure is  that the weak
``characteristic''  background is ignored. We discuss these issues
further in Appendix~\ref{app:TRS_analysis} where the TRS and MASME
analysis methods are carefully compared.

\subsection{Loop modeling \label{loop_modeling_analysis}}

One of the major goals of our flare analysis is to derive sizes of
coronal structures responsible for the COUP flares. We employ the
time-dependent hydrodynamic model of \citet{Reale97} for a single
coronal magnetic loop.  Reale et al.\  establish a formula for
estimating loop size taking accounting for the possibility of
prolonged heating during the decay phase. They find
\begin{equation}
L = \frac{\tau_{d} \sqrt{T_{pk}}}{3.7 \times 10^{-4} F(\zeta)}
\end{equation}
where $L$ is the half-length of the loop (cm), $\tau_{d}$ is the
flare decay e-folding timescale (sec), and $T_{pk}$ is the flare
maximum temperature at the loop apex (K). $F(\zeta)$ is a
correction factor for prolonged heating that is a function of the
slope $\zeta$ of the trajectory in the temperature-density
diagram. In practice, $F(\zeta)$ and $T_{pk}$ must be calibrated
for each X-ray telescope and detector; the slope $\zeta$ is
usually measured in the $\log T - \log (EM^{1/2})$ plane where
$EM$ is the changing emission measure and $EM^{1/2}$ is used as a
proxy for the plasma density. We reproduce calibration formulas
for $F(\zeta)$ and $T_{pk}$ derived for $Chandra$-ACIS in
\citet{Favata05}
\begin{equation}
F(\zeta) = \frac{0.63}{\zeta-0.32} + 1.41
\end{equation}
and
\begin{equation}
T_{pk} = 0.068 \times T_{obs}^{1.2}
\end{equation}
where $T_{pk}$ is a temperature at the loop apex (K), and
$T_{obs}$ is an ``average'' loop temperature (K) obtained from our
{\it Chandra}-ACIS data.

This model is simplistic in a number of ways.  It assumes the
plasma has a uniform density with unity filling factor confined
within a single semicircular loop of uniform cross-section and
10:1 length:radius shape. Furthermore, the model assumes that this
geometry remains unaltered during the flare, that energy is
efficiently transported along magnetic field lines of the loop,
and that there is continuous  energy balance between the loop
heating and the thermal conduction and radiative losses. Despite
these limitations, the \citet{Reale97} model has been applied
successfully  to a variety of solar and stellar flares
\citep[e.g.][]{Crespo-Chacon07,Testa07,Franciosini07,Giardino07},
including the COUP flares studied by \citet{Favata05}.  It has the
advantage over earlier and simpler cooling loop models
\citep[e.g.][]{Rosner78,Serio91} which neglect reheating during
the decay phase and thereby tend to overestimate loop sizes.

\subsection{Flare Characteristics \label{derived_quantities_analysis}}

Empirical and derived quantities obtained in our MASME flare
analysis are reported in Tables \ref{tbl_flare_prop1} and
\ref{tbl_flare_prop2}. We explain them in this section and
illustrate them in detail for one flare in Figure
\ref{fig_flare_example}. The electronic edition of this paper
contains a figure set  similar to Figure \ref{fig_flare_example}
for our full sample (216 flares). Many sources show more than one
bright flare during the 13-day COUP observation. In two sources
(COUP \#682 and 1463) we have analyzed four bright flares, ten
sources have three flares; and forty three sources have two
flares.  The $FlareNo$ quantity in conjunction with the $Source$
number in Tables \ref{tbl_flare_prop1} and \ref{tbl_flare_prop2}
provides a unique identifier for each of the 216 flares.

Several reported quantities are associated with the characteristic
level between flares.   Quantities t$_{char1}$ and t$_{char2}$
represent the start and stop times (offsets from the beginning of
the COUP observation) of the ``characteristic'' lightcurve block
with the lowest count rate. These intervals were defined visually
using the interactive graphical tool $function\_1d$ in the $TARA$
package; the duration of this block is always $>20$~ks.  The
median count rate in this interval is CR$_{char}$.  These
quantities are labeled in blue in Figure
\ref{fig_flare_example}$a$.

The flare events are defined by the quantities t$_{flare1}$ and
t$_{flare2}$ (labeled in red in Figures \ref{fig_flare_example}$a$
and $b$) giving start and stop times for the flare, with the peak
count rate given by CR$_{flare,pk}$. Both CR$_{char}$ and
CR$_{flare,pk}$ are obtained from the binned histogram data (black
points in Figures \ref{fig_flare_example}$a$ and $b$), and not
from the adaptively smoothed version of the lightcurve (dashed
blue curve in Figure \ref{fig_flare_example}$b$). Following the
recommendation of \citet{Reale97}, we choose the flare stop time
as the moment at which the flare count rate has decreased to $\sim
1/10$ of the peak value.  However, often an earlier t$_{flare2}$
was chosen because a COUP observational gap or minor flare fell
near the end of the studied flare.

Quantity $Bgf$ represents the number of background counts expected
in the flare time interval, obtained by scaling the source's
background from \citet{Getman05} to the duration of the flare.
With a few exceptions, this correction is very small ($Bgf/NCf <
0.02$). We therefore have chosen to ignore the background in
estimating the median energy $MEf$ during the flare.

Quantities $\tau_{r}$, $\tau_{d1}$, and $\tau_{d2}$ give $1/e$
rise time and decay times determined from a least-squares fit of
the count rate histogram.  This is illustrated by the red and
green curves on Figure \ref{fig_flare_example}$b$.   These are
fits of the exponential function ${\rm CR}={\rm CR}_{flare,pk}
\times \exp[({\rm t}-{\rm t}_{CR,pk})/\tau]$ where t$_{CR,pk}$ is
the time of the peak count rate, and $\tau$ is $-\tau_{d1,2}$ or
$+\tau_{r}$. Decays of many COUP flares deviate from a pure
exponential shape. This is not unusual --- \citet{Osten99} found
that 8 of 33 flares from RS CVn binaries observed with the Extreme
Ultraviolet Explorer showed changing exponential decays.  Complex,
often bumpy, decays arise from both reheating of a single loop and
the illumination of associated loops. These issues are discussed
in the next section. For COUP X-ray flares, we introduce two decay
e-folding timescales, $\tau_{d1}$ for the initial purely
exponential decay phase lasting at least to the time point
t$_{exp}$, and $\tau_{d2}$ for the e-folding decay timescale
averaged over the entire decay phase from t$_{CR,pk}$ until
t$_{flare2}$.  As the data become less consistent with a single
exponential decay model, the difference between $\tau_{d1}$ and
$\tau_{d2}$ increases.

The adaptively smoothed count rate and median energy time series
are shown as the blue dashed curve in Figure
\ref{fig_flare_example}$b$ and the black symbols in Figure
\ref{fig_flare_example}$c$.  These are converted into plasma
temperature (Figure \ref{fig_flare_example}$d$) and flare emission
measure (Figure \ref{fig_flare_example}$e$) time series using
calibration simulations described in \S
\ref{sliding_kernel_method_analysis}. Information on maximum
values of flare X-ray luminosity ($\log L_{X,pk}$), observed
temperature (${\rm T}_{obs,pk}$), emission measure (${\rm
EM}_{pk}$), and corresponding time points, such as t$_{T,pk}$ and
t$_{EM,pk}$ is then easily obtained.

Considering that the proper decay phase begins at the time of the
density peak, not count rate or temperature peak (at least for an
impulsive flare event), \citet{Reale07} encourages using the
temperature at the density peak rather than the brightness peak
when determine the loop length. We thus introduce the quantity
T$_{EM,pk}$, the plasma temperature at the peak of the emission
measure. Cross symbols in Figure \ref{fig_flare_example}$c$-$f$
mark the rise phase, the time period prior to the moment when
emission measure reaches its maximum value, whereas  circle
symbols mark the decay phase of the flare.

Several remaining obtained quantities listed in Tables
\ref{tbl_flare_prop1} and \ref{tbl_flare_prop2} and shown in
Figure \ref{fig_flare_example} remain to be discussed. Quantity
t$_{sep}$ is the time point separating the two distinct decay
phases when a simple exponential decay is not present.  Quantities
$\zeta_1$, $\zeta_2$, and $\zeta_3$ are the inferred slopes of the
trajectory in the $\log T - \log(EM^{1/2})$ diagram.  The
morphological flare type is discussed in \S
\ref{morphology_analysis}. Quantities L$_1$, L$_2$, and L$_{lim}$
are the lengths of the flaring structures inferred from equation
(1).  Flag$_1$ provides details on a number of decay phases and
indicates the ``dominant'' slope used in estimation of loop sizes.

\subsection{Flare Morphology and Classification \label{morphology_analysis}}

In order to obtain optimal results within the framework of the
\citet{Reale97} single loop model, the X-ray lightcurve should not
present significant deviations from a pure exponential decay, and
the decay trajectory in the $\log T - \log (EM^{1/2})$ plane
should be linear. Strong deviations from an exponential decay and
the absence of a single linear trajectory are indications of more
complex events such as multiple coronal loops (e.g. sequential
reconnection in a loop arcade as in a 2-ribbon solar flare) and/or
complex heating sequences within a single loop
\citep[e.g.][]{Reale04,Reale07}.

However, even in the case of complex flares it is appropriate to
apply the single loop approach to a lightcurve segment if there is
an indication for the presence of a single dominant loop.
\citet{Reale07} argues that when multiple loop structures are
involved in the flare, the rise and early decay are often
dominated by a single primary loop. This is supported by detailed
time-dependent loop modeling of the ``bumpy'' decay flare observed
with XMM-Newton on Proxima Centauri \citep{Reale04}. Here it was
found that the X-ray emission from the primary single loop
dominated the initial phases of the flare, but later was
overlapped by the emission from an arcade of several loops of
similar size as the primary loop.

Similar complex, multiple-component flares have been seen in the
contemporary Sun, including the well-studied Bastille Day flare in
July 2000 \citep{Aschwanden01} and a variety of bumpy flares seen
with the GOES and TRACE satellites.  They are seen in both
2-ribbon flares and compact flares\footnote{Examples of ``bumpy''
solar flares seen in soft X-rays by GOES~8, and EUV light by
TRACE, can be found in the TRACE Flare Catalog
(\url{http://hea-www.harvard.edu/trace/flare\_catalog/index.html}).
See, for example,  the 16 Jan 05 (22:03:00UT, M2.4) and 08 Sep 05
(20:29:00UT, M2.1) 2-ribbon flares and the 14 Nov 05 (22:00:00UT,
M1.0) and 02 Dec 05 (10:12:00UT, M7.8) compact flares.}.
\citet{Reale04} speculate that this is a common pattern in large
solar and stellar magnetic reconnection.

We have adapted these ideas, together with our careful examination
of the X-ray morphology of the 216 COUP flares, into a
classification scheme consisting of four well-defined groups and
two groups to gather flares which cannot be clearly characterized.
In order of their morphological complexity, these classes are:
\begin{description}

\item[Typical flares (typ)]  These have simple X-ray lightcurves with
rapid monotonic rise and generally slower monotonic decay.  This
is the most populous class with 84 of the 216 COUP flares. Seven
of them have symmetrical lightcurve shape with similar rise and
decay timescales.

\item[Step flares (stp)]  These look like a typical flare
but with a shoulder or bump overlayed on its decay phase. These
morphologies are commonly seen on the Sun and are due either to a
reheating event or to a triggered reconnection in a nearby
magnetic loop. Thirty-eight COUP flares fall into this category.
The ``bumpy'' decay flare from Proxima Centauri \citep{Reale04} is
a good example of this class.

\item[Double flares (dbl)] Double flares look like two overlapped
typical flares, or a typical flare with a bump on its rise phase.
Eight flares clearly show double peaks in appearance.

\item[Slow rise, top flat flares (srtf$=$srf)]  These are more complex
events where variations appear to occur more slowly than in most
flares. They have slow rises, long duration peaks, and/or very
long decays. One of these (a T-Tauri star in the Orion B cloud)
was discussed by \citet{Grosso04} without a clear interpretation,
and a few (COUP \#43, 597, 1384) appear in \citet[][see their
Figure 9]{Favata05}. In our study, twenty COUP flares are in this
class.

\item[Incomplete (inc)] These are 42 flares with their {\it Chandra}
lightcurves severely interrupted by the COUP observational gaps.

\item[Others (oth)] These are 24 flares with poorly defined
shapes generally due to their low statistics.

\end{description}

The four well-defined classes are illustrated in Figures
\ref{fig_morphology_typical}-\ref{fig_morphology_dbl} along with
their inferred trajectories on the $\log T - \log (EM^{1/2})$
plane. For about 56\% (34\%) out of 84 (38) typical (step) flares,
trajectories can be approximated by a single slope. In these
cases, the derived loop sizes are most reliable. In many of the
remaining cases, two slopes can be discerned in the $\log T - \log
(EM^{1/2})$ trajectories, while in some the patterns are too
complex to interpret.  Flare classifications based on lightcurve
morphology may not be correct in all cases; for example,  a few
flares with simple lightcurve decays classified as typical flares
show double-peaked temperature evolutions.  These effects are
included in the loop reheating model (\S
\ref{loop_modeling_analysis}) and thus the derived loop sizes, but
such flares have not been reclassified as double-peaked flares.

Flag$_1$ in Table \ref{tbl_flare_prop2} encodes the number of
inferred dominant decay slopes of $\log T - \log (EM^{1/2})$
trajectories; values of $01$ ($=\zeta_1$), $02$ ($=\zeta_2$) and
$03$ ($=\zeta_3$) refer to a single dominant slope, while $12$
($=\zeta_1$,$\zeta_2$), $13$ ($=\zeta_1$,$\zeta_3$) and $23$
($=\zeta_2$,$\zeta_3$) refer to a double slope. In cases with a
single slope, the slope value used is generally the average slope
$\zeta_1$ (green solid line in Figure \ref{fig_flare_example}$f$)
derived within the the whole flare time interval.  In double slope
cases, the two slopes are generally described by $\zeta_2$ (slope
within the initial decay; red dotted line in Figure
\ref{fig_flare_example}$f$) and $\zeta_3$ (slope within the late
decay; orange dashed line in Figure \ref{fig_flare_example}$f$)
separated in time by the point t$_{sep}$ (given in
Table~\ref{tbl_flare_prop1}).

While we do not know of a similar flare classification effort in
other solar or stellar studies, a similar range of flare behaviors
is seen in the contemporary Sun.  Analysis of Solar Maximum
Mission observations of 66 solar flares by  \citet{Sylwester93}
showed 26\% of flares have $\log T - \log (EM^{1/2})$ trajectories
with two or more branches while the majority show single-branched
trajectories.

\subsection{Flare Loop Sizes \label{sizes_analysis}}

Due to the complexity of flare behavior in our sample (described
above, shown in Figure \ref{fig_flare_example} and in the
electronic atlas), it seems reasonable to present a range of
possible loop sizes rather than a single size value for a given
flare. The range of sizes is derived with the assumption for the
presence of dominant coronal structures even during the complex
flare events. This approach should be suitable for the goals of
the current work, which is oriented towards examination of
statistical relationships between flare behaviors, magnetic loop
sizes, global star properties, protoplanetary disks and accretion.

The loop size range ($[\rm{L}1:\rm{L}2]$, reported in Table
\ref{tbl_flare_prop2}) is estimated through equations (1)$-$(3) in
\S \ref{loop_modeling_analysis} using the combination of both
decay timescales (initial, $\tau_{d1}$, and averaged over the
entire decay phase, $\tau_{d2}$) with the decay slopes indicated
by Flag$_1$. If a single dominant slope is indicated by Flag$_1$,
then two loop size estimates emerge after applying both decay
timescales with the combination of that single slope.  If a double
slope is indicated by Flag$_1$, then two loop size estimates
emerge in the following ways:  for Flag$_1=23$, we combine
$\tau_{d1}$ with $\zeta_2$ and $\tau_{d2}$ with $\zeta_3$; for
Flag$_1=13$, we combine $\tau_{d1}$ with $\zeta_1$ and $\tau_{d2}$
with $\zeta_3$; and for Flag$_1=12$, we combine $\tau_{d1}$ with
$\zeta_2$ and $\tau_{d2}$ with $\zeta_1$. $T_{obs}$ in formula (3)
is the observed {\it Chandra}-ACIS flare temperature at the peak
of the emission measure ($T_{EM,pk}$ from Table
\ref{tbl_flare_prop2}).

L$_{lim}$, the limiting maximum loop length assuming a freely
decaying loop with no sustained heating, is computed using
$\tau_{d2}$ in equation (14) of \citet{Serio91}.  This estimate is
based on the older cooling loop theory of \citet{Rosner78} and may
over-estimate loop lengths if reheating or triggered events extend
the decay phase.

We compared our methodology with the well-studied case of the
August 2001 flare on the dMe star Proxima Centauri observed with
XMM-Newton \citep{Reale04}.  This flare has a ``bumpy'' decay
similar to our ``step'' (stp) flares.  Using our procedures, we
obtain $\zeta_1 = 0.6$, $\zeta_2 = 0.7$, $\zeta_3 = 0.4$,
$\tau_{d1} = 1.7$~ks, $\tau_{d2}=3.3$~ks, t$_{sep}=4.7$~ks,
Flag$_1=01$.  The inferred loop size range is L$_1 =
5\times10^{9}$~cm and L$_2=1\times10^{10}$~cm. \citet{Reale04}
obtained a value of $1\times10^{10}$ from a detailed
time-dependent hydrodynamic model of the flare evolution. We
conclude that our procedures are consistent with other analyses of
similar complicated flares.

\subsection{Collected Source Properties \label{source_properties}}

Properties of the 161 COUP stars hosting our flare sample are
presented in Table \ref{tbl_source_prop}. These stellar properties
will be extensively used in Paper II. Many of the quantities have
been taken from the COUP tables of \citet{Getman05} including:
time-integrated net counts, background counts, median energy,
column density. $JHK_s$ magnitudes obtained from the merged
2MASS-VLT catalog with a flag indicating quality of the 2MASS
photometry. Optical data include the 8542~\AA\ Ca~{\tiny II} line
strength as an accretion indicator, and stellar mass and radius
estimated from position in the Hertzsprung-Russell diagram. See
Getman et al. for full details and references for these
quantities. Stellar rotational periods are taken from the extended
rotational data collection of \citet{Flaccomio05}. For stars with
known masses ($M$) and rotational periods ($P$) we calculate the
Keplerian corotation radius where centrifugal force balances
gravity from $R_{cor} = (G \times M \times P^2/4 \times
\pi^2)^{1/3}$.

The near-infrared (NIR) color excess $\Delta(H-K_s)$, an indicator
of dusty inner circumstellar disks, is calculated here as an
excess from the rightmost reddening vector on the $J-H$ vs.
$H-K_s$ color-color diagram originating at $\sim 0.1$~M$_{\odot}$
assuming an age of 1~Myr and using PMS models of \citet{Siess00}.
$\Delta(H-K_s)$ is given only for sources with reliable 2MASS
photometry (${\rm Fl}=$AAA000).

A more sensitive measure of disk dust is derived from mid-infrared
(MIR) images of the Orion region in the 3.6 and 4.5~$\mu$m IRAC
bands using the  $Spitzer$ archived data ($Spitzer$ observing
program \#50; G. Fazio, PI). For each band, the $3 \times 4$
shallow maps, each with 24 dither pointings, processed by the
Spitzer Science Center pipeline (ver. S14.0.0) were combined into
one mosaic with pixel size 0.86$\arcsec$ using IDL software
developed by R. Gutermuth of the IRAC instrumental team.  Aperture
photometry was performed with the routine {\it aper.pro} in the
IDLPHOT package\footnote{IDLPHOT is available at
\url{http://idlastro.gsfc.nasa.gov/ftp/pro/idlphot/aaareadme.txt}.}
for isolated COUP sources using an aperture radius of 4 pixels
with a $4-8$ pixel annulus for sky subtraction.  For these
aperture and sky regions, we adopted zero-point magnitude values
of 19.490 and 18.751~mag, for the 3.6 and 4.5~$\mu$m bands,
respectively (K. Luhman, private communication).   High
signal-to-noise photometric results in both bands are presented
for 112 stars in Table \ref{tbl_source_prop}. The majority of
stars without MIR photometry are those located in the outer
regions of the Orion Nebula where [3.6] and [4.5] images do not
overlap. For several sources located near the BN/KL region, MIR
photometry is unreliable due to strong nebular contamination in
infrared images, and is omitted.

\section{RESULTS \label{results}}

\subsection{Global flare properties  \label{global_properties}}

Figure \ref{fig_global_props} summarizes some of the inferred
physical properties of the COUP flares in the form of univariate
cumulative distributions differentiated by the flare morphological
classes defined in \S \ref{morphology_analysis}.  The panels plot
distributions of several quantities relating to flare rise and
decay timescales, peak luminosities, and peak temperatures.

Panel $a$ of Figure \ref{fig_global_props} shows that 90\% of
these powerful COUP flares have rise times between 1 hour and 1
day ($3-100$~ks).  ``Typical'' (typ) and ``step'' (stp) flares
have similar rise times with medians around $\sim 3$ hours
(10~ks). In contrast, as expected from their class definition,
``slow-rise-top-flat'' (srtf) flares have much longer rise
timescales with median around 12 hours (44~ks)\footnote{We caution
that our procedure for estimating the e-folding rise timescales
for COUP ``srtf'' flares, in which the rise phase is defined as
the time interval between the flare start and the time point with
a maximum observed count rate, may give longer timsecales than
those derived in other works. For example, \citet{Grosso04} report
$\tau_{rise} \sim 2$~hours (7.2~ks) for the slow-rise flare in
LkH$\alpha$~312 by restricting the rise to the period of very
rapid count rate change. For this flare, our procedure would give
$\tau_{rise} \sim 5$ hours ($\sim 20$~ks).}. It is unclear whether
this slow rise is due to multiple heating events of a single
magnetic loop, or a sequence of triggered reconnection events in a
loop arcade.  In contrast to all of these PMS flares, a rise time
of the soft X-ray emission in a typical solar flare is typically
only a few minutes \citep[e.g.][]{Priest02} because solar
chromospheric material fills the smaller solar flare volumes more
quickly. \citet{Franciosini07} provide statistics on rise and
decay times for Taurus flares from the XEST study.

Panel $b$ shows that the decay e-folding timescales derived over
the whole flare interval, ($\tau_{d2}$), ranging from 3 hours to
1.5 days (10~ks to 150~ks) for 90\% of the COUP flares.  The
median of $\tau_{d2} \sim 6$ hours (22~ks) for ``typical'' flares
is much shorter than that of all other morphological flare classes
which have median decay times around 12~hours ($\sim 45$~ks). The
$\tau_{d1}$ decay timescales derived over the initial exponential
decay phase $[{\rm t}_{CR,pk}-{\rm t}_{exp}]$ are on average a
factor of $1.2$ shorter than $\tau_{d2}$, again with decays
systematically shorter in ``typical'' flares than other classes
(panels $c$ and $e$).   The decay timescales in typical solar
flares are even shorter or comparable with this, and are of
$10$~min to a few hours \citep[e.g.][]{Reale97,Priest02}. The
ratios of rise and decay timescales serve as good quantitative
discriminators between major different morphological flare
classes.  The median ratio $\tau_{d2}/\tau_{raise}$ is 1.2, 2.6,
and 4.3~ks for ``srtf'', ``typical'', and ``step'' flares,
respectively (panel $d$).  The median of $\tau_{d2}/\tau_{d1}$ for
``step'' flares of 1.7 is much higher than the values of 1.1-1.2
for the ``typical'', ``srtf'' and other classes (panel $e$).

Panel $f$ of Figure \ref{fig_global_props} indicates that peak
flare X-ray luminosities in the$[0.5-8.0]$~keV  band for our flare
sample span $\log L_{x,pk} \simeq 30.6$ to 32.3~erg~s$^{-1}$ (90\%
range) with $10\%$ of flares reaching luminosities $\log L_{x,pk}
\ga 32$~erg/s. The strongest flare, flare~\#1 from COUP~\#1462,
has peak luminosity of 32.9~erg s$^{-1}$; it is also one of the
hottest flares ($T_{obs,pk} \gg 200$~MK). While the lower end of
this distribution has no scientific importance due to our
selection criteria, it is relevant that no significant luminosity
differences are seen between different flare morphological
classes, except for a possible indication that ``srtf'' flares do
not achieve the highest peak luminosities. Similarly, no
statistical differences in temperature distributions are seen
(panel $g$). Temperatures span 20 to $\sim$500~MK (90\% range)
with median around 63~MK.  Recall that individual values of
temperatures above $\sim 200$ MK are quite uncertain (Figures
\ref{fig_kt_vs_mede} and \ref{fig_MASME_vs_TRS_general}b).

These results thus characterize the global properties of PMS
flares with peak X-ray luminosities between 31 and
33~erg~s$^{-1}$.   The temperatures and luminosities of the
brightest COUP flares are similar to the most luminous X-ray
flares ever recorded in young stellar objects: V773 Tau in the
Taurus clouds with $L_{pk} \ga 10^{33}$~erg~s$^{-1}$ and $T_{pk}
\ga 100$~MK \citep{Tsuboi98}, YLW~16A in the $\rho$ Ophiuchi Cloud
with $L_{pk} \sim 2 \times 10^{32}$~erg~s$^{-1}$ and $T_{pk} \sim
140$~MK \citep{Imanishi01}, LkH$\alpha$~312 in the M~78 reflection
nebula of Orion with $L_{pk} \simeq 10^{32}$~erg~s$^{-1}$ and
$T_{pk} \simeq 90$~MK \citep{Grosso04}, and star \#294 in the
Cep~OB3b region with $L_{pk} \simeq 2 \times 10^{32}$~erg~s$^{-1}$
and $T_{pk} \gg 100$~MK \citep{Getman06}. Recall that even the
most powerful solar flares today rarely exceed 28.5~erg~s$^{-1}$
in the Chandra band.

While the differences between rise and decay timescales between
our morphological classes mostly reflect our class definitions,
the similarities in $L_{X,pk}$ and $T_{obs,pk}$ distributions
among classes (Figure \ref{fig_global_props}f and g) are real
results. They demonstrate that differences in flare morphology --
particularly the difference between typical flares with fast-rises
and exponential-decays and atypical flares with slow-rises and
flat-tops -- are not reflected in large differences in luminosity
or temperature.  This suggests that all types of flare
morphologies arise from similar magnetic reconnection mechanisms.

\subsection{Comparison with flares on older stars \label{Achwanden07_comp_section}}

\subsubsection{Comparison with general solar-stellar scaling law \label{solar_stellar_law}}

From their compilation and comparison of comprehensive data sets
of solar and stellar\footnote{Stellar flare parameters for PMS and
older active stars were taken from the compilation of
\citet{Gudel04}.} flares, \citet{Aschwanden08} find a common
scaling law between peak flare emission measure ($EM_{pk}$) and
peak flare plasma temperature ($T_{obs,pk}$), $EM_{pk} \propto
T_{obs,pk}^{4.7}$, but with stellar emission measures $\sim 250$
times higher than that of solar flares. They also find that solar
and stellar flare durations, $\tau_f$, follow the trend $\tau_f
\propto T_{obs,pk}^{0.9}$.  For flare loop lengths, $L$, they find
the trend  $L \propto T_{obs,pk}^{0.9}$ for solar flares only.
Here we relate our 216 COUP flares to these scaling laws of
solar-stellar flares.

Figure \ref{fig_compare_older_stars1} compares the
$EM_{pk}-T_{obs,pk}$ relation between COUP flares (black circles),
stellar flares of \citet{Gudel04} (grey boxes), and solar-stellar
trends obtained by \citet{Aschwanden08} (dashed and dotted lines).
The COUP flares represent the most luminous flare events ever
detected from stellar objects, occupying the highest range in
stellar emission measures,  $53.5 < \log(EM_{pk}) < 56$ cm$^{-3}$.
The lower boundary corresponds to our selection cut of $NC>4000$
source counts (\S \ref{source_sample_analysis}).

In agreement with the scaling-law of \citet{Aschwanden08}, about
$60\%$ of the COUP flares are located within the $67\%$ range of
flares from older stars (dashed lines).  But the remaining stars
do not follow the scaling-law of \citet{Aschwanden08}: $34\%$ (73
out of 216) of the COUP flares with unexpected super-hot plasma
temperatures of $T_{obs,pk} \ga 100$~MK strongly deviate from the
expected  $EM_{pk} \propto T_{obs,pk}^{4.7}$ trend.  These
deviations may not be restricted to COUP flares; a careful
examination of the stellar flare sample of \citet{Gudel04} shows
that 10 out of their 11 flares (grey boxes) with $T_{obs,pk} \ga
100$~MK are located below the regression line of
\citet{Aschwanden08}, in agreement with the COUP super-hot flares.
Considered by themselves, the COUP flares follow a much shallower
scaling law, $EM_{pk} \propto T_{obs,pk}^{0.4}$.  The statistical
significance of this shallow correlation is strong: COUP flares
with $T_{obs,pk} < 100$~MK have  median $\log(EM_{pk}) =
54.09$~cm$^{-3}$ while super-hot flares with $T_{obs,pk} \ga
100$~MK with median $\log(EM_{pk}) = 54.45$~cm$^{-3}$ with
two-sample probability of similar $EM$ distributions $P_{KS} <<
10^{-4}$.

Figure \ref{fig_compare_older_stars2} compares flare duration
$\tau_f$ to  peak plasma temperature $T_{obs,pk}$ for our COUP
flares (black circles), stellar flares compiled by \citet{Gudel04}
(grey boxes) and the regression fits for solar-stellar flares
obtained by \citet{Aschwanden08} (dotted lines). Here we use our
longest decay timescale $\tau_{d2}$ to correspond with their flare
durations $\tau_f$. Flare durations for solar LDEs (long decay
events, see \S \ref{solar_LDEs}) are $\tau_f=(t_{end}-t_{start})$
on their GOES lightcurves which are, on average, 3 times longer
than their e-folding decay timescales.

First, we note that the remarkably long duration (13 days) of the
COUP observation provided a unique opportunity to capture stellar
flares with decay timescales longer than a day. As in the $EM_{pk}
- T_{obs,pk}$ plot, the COUP flares do not exhibit the reported
solar-stellar relationship, $\tau_f \propto T_{obs,pk}^{0.9}$.
While some COUP flares studied here are consistent with the
relation, the majority deviate strongly with peak plasma
temperatures either too cool or too hot.   For COUP flares, we
find a weak trend in the other direction, $\tau_{d2} \propto
T_{obs,pk}^{-0.2}$. Similar results are obtained using our flare
rise timescale $\tau_{r}$ and initial flare decay timescale
$\tau_{d1}$, and the trends are confirmed with Kolmogorov-Smirnov
tests\footnote{KS results indicate significant differences between
distributions of $\tau_{r}$ ($P_{KS}=0.004$) with median values of
$\tau_{r} = 15.8$~ks and $\tau_{r} = 9.7$~ks for $T_{obs,pk} <
100$~MK and $T_{obs,pk} \ga 100$~MK flares, respectively; and
$\tau_{d1}$ ($P_{KS}=0.001$) with median values of $\tau_{d1} =
30.1$~ks and $\tau_{d1} = 21.5$~ks for $T_{obs,pk} < 100$~MK and
$T_{obs,pk} \ga 100$~MK flares, respectively.}.  The median
duration for flares with $T_{obs,pk} < 100$~MK is $\tau_{d2} =
45$~ks compared to 25~ks for super-hot flares with $T_{obs,pk} \ga
100$~MK. However, we do not analyze here many hundreds of weaker
and shorter COUP flares that could potentially compensate some of
these deviations.

Although stellar and COUP flare loop lengths are derived from
cooling models rather than directly from the observations, it is
useful to compare COUP, stellar, and solar length distributions.
Figure \ref{fig_compare_older_stars3}$a$ shows the COUP flares
(black circles), stellar flares (grey boxes), and locus of solar
flares (dotted lines) in the $L$-$T_{obs,pk}$ diagram.   For COUP
flares, $L$ is the mean value of the inferred $[L_1-L_2]$ range.
COUP loop sizes range from $< 10^{11}$~cm to $\ga 10^{12}$~cm, a
range of very large loops previously occupied by only dozen
stellar flares. These loop sizes are $10^2-10^3$ larger than seen
in solar flares.  Figure \ref{fig_compare_older_stars3}$b$ shows
that the variables $L$ and $L/R_*$ are roughly equivalent; COUP
flare loop lengths range from $\sim 0.2$ to $\ga 10$ stellar
radii. COUP loop lengths slowly increase with peak flare
temperature roughly as $L \propto T_{obs,pk}^{(0.4-0.5)}$,
consistent with the slope seen in solar flares but offset to much
larger lengths at a given temperature. This trend, however,
emerges from the assumptions underlying our loop modeling (\S
\ref{loop_modeling_analysis}, equations (1) and (3)) and thus may
not be physically significant.

We thus find that flares in our COUP sample are indeed among the
most luminous, hottest, largest (in term of loop length) and
long-lived of any X-ray flares known. Comparison with the
stellar-solar scaling laws of \citet{Aschwanden08} provides
non-trivial results. One-third of the COUP sample have super-hot
temperatures of $T_{obs,pk} > 100$~MK.  These do not fit the
strong $EM_{pk} \propto T_{obs,pk}^{4.7}$ solar-stellar scaling
law; rather, in the COUP regime, temperatures can increase several
fold above 100~MK with only slight increase in plasma emission
measures. This may be related to the low efficiency of filling up
very large coronal loops of PMS stars with hot plasma from
chromospheric evaporation. COUP flares also do not follow the
solar-stellar correlation between flare duration and peak plasma
temperature; indeed, we find  a slight anti-trend where hotter
COUP flares are on average slightly shorter in duration than
cooler flares. However, a previously reported correlation between
flare loop length and peak plasma temperature is present, though
offset to longer loops than seen in solar and older stellar
flares. Further analysis of COUP super-hot flares is presented in
Paper~II.

Some form of saturation of previously reported relationships is
clearly present. Unlike solar flares which arise in small-scale
multi-polar fields, COUP flaring magnetic structures are typically
larger than the radii of their host stars.   We turn now to
possible relationships with a rare class of long duration solar
flares which are also believed to be associated with large scale
coronal structures.

\subsubsection{Solar long decay flare events \label{solar_LDEs}}

Giant X-ray arches and streamers with altitudes reaching up to
several hundred thousand kilometers produced during some solar
flares have been observed by the {\it Skylab}, SMM and {\it
Yohkoh} space observatories \citep[][and references
therein]{Hiei94,Svestka95,Farnik96,Svestka97,Hiei97}. Such events
were originally called long decay events
\citep[LDEs,][]{Kahler77}. Corresponding X-ray lightcurves of
solar LDEs often exhibit flares lasting from a few hours to a day,
similar to flare durations in our sample. Figure
\ref{fig_goes_lde} exemplifies the GOES lightcurves of long solar
flares associated with giant X-ray arches. The origin of solar
X-ray giant arches and streamers is not well understood. A popular
model explains that the powerful prominence eruption or coronal
mass ejection, often (but not always) associated with an impulsive
flare \citep{Priest02}, expands into an overlying large-scale
magnetic field with the subsequent reconnection of magnetic lines
through a vertical current sheet and formation of a system of
giant closed loops in a shape of arches and streamers
\citep{Sturrock66,Kopp76,Forbes96}.

Flare parameters for a number of solar LDEs are shown in Figures
\ref{fig_compare_older_stars1}-\ref{fig_compare_older_stars3}
(grey diamonds)\footnote{Flare parameters are presented for the
following solar LDEs: giant arches observed on 2 Nov 1991, 4 Nov
1991, 21 Feb 1992, and 15 Mar 1993; X-ray streamers seen on 24 Jan
1992 and 28 Oct 1992. Peak flare emission measures, plasma
temperatures, and maximum altitudes are taken from
\citet{Svestka95,Svestka97,Getman00}. Flare durations are
$\tau_f=(t_{end}-t_{start})$ in their GOES lightcurves.}. Peak
emission measures of solar LDEs vary from $\log(EM_{pk}) \sim
44$~cm$^{-3}$ for X-ray streamers to $\log(EM_{pk}) \sim
50$~cm$^{-3}$ for giant arches, with flare durations $\tau_f >
10$~ks and altitudes of $ L \ga 1 \times 10^{11}$~cm. Both
durations and altitudes are well above loci of typical solar
flares.  In addition, several solar flare limb events observed on
$Skylab$ with inferred durations of $>12000$~seconds, temperatures
of $>7-8$~MK, and altitudes of $>3-4 \times 10^{10}$~cm, which are
called Class~II events \citep{Pallavicini77}, are also located far
the usual solar flare locus \citep[see diamond symbols in Figure~4
of][]{Aschwanden08}. These may represent a sub-class of solar
LDEs.

With their extremely long flare durations, often hot temperatures,
and associated large-scale flaring structures, LDEs are thus
outliers from the solar flare scaling laws, similar in many
respects to the COUP flares analyzed here. It is reasonable to
propose that the majority of the flares in our COUP flare sample
are enhanced analogs of these solar LDEs. Concerns regarding
possible effects of centrifugal forces and magnetic field
confinement on the COUP PMS flares are discussed in Paper II.

\section{CONCLUSIONS \label{conclusion_section}}

We analyze 216 bright X-ray flares from the Chandra Orion
Ultradeep Project which provides the longest nearly-continuous
observation of a rich PMS stellar cluster in the X-ray band. Our
effort is based on a new spectral analysis technique (MASME) that
avoids nonlinear parametric modeling and is more sensitive than
standard methods.  Flare loop parameters are derived from the
well-established flare plasma model of \citet{Reale97}.

We thus emerge with the largest dataset of PMS flares, or indeed
stellar flares at any stellar age, with peak luminosities in the
range  $31< \log L_{X,pk}< 33$ erg~s$^{-1}$, several orders of
magnitude more powerful than any solar flare. For each flare we
provide a catalog of $>30$ derived flare properties including
inferred sizes of associated coronal loops and flare morphological
classes. We give an electronic atlas with flare lightcurves,
temporal evolution plots of X-ray median energy, plasma
temperature, emission measure, and derived temperature density
diagram. This collection of empirical and model-dependent
information can serve as a valuable testbed for stellar flare
models.

The powerful COUP flares studied here have rise timescales ranging
from 1~hour to 1~day and decay timescales ranging from a few hours
to 1.5 days.  An important empirical result is that peak plasma
temperatures are often 100~MK, in some cases $> 200$~MK. These
temperatures are derived from a robust calibration of median
energies; traditional time-resolve spectroscopy often does not
have the time resolution to detect this brief super-hot phase. No
significant differences in peak flare luminosity or temperature
distributions are found among the wide range of morphological
flare classes: typical fast-rise exponential-decays, step decays,
double peaks, and slow-rise flat-top.  This suggests that all
flare types arise from similar underlying magnetic reconnection
mechanisms and geometries.

Comparison of the COUP flare properties with the general
solar-stellar scaling laws of \citet{Aschwanden08} presents
surprising results. Our flares do not follow the solar-stellar
trend between plasma peak emission measure and temperature,
$EM_{pk} \propto T_{obs,pk}^{4.7}$. The trend between flare
duration and peak temperature is also absent. Super-hot COUP
flares are found to be brighter but shorter in duration than
cooler COUP flares. This is further developed in Paper II.

Compared to non-PMS systems, the inferred sizes of COUP flaring
structures are remarkably large, ranging widely from $L=0.5$ to
$10$~R$_{\star}$. These large flaring structures must be
associated with large-scale stellar magnetic fields. Rare long
decay solar events associated with the largest known X-ray
emitting structures are possible solar analogs to these COUP
flares.

Our flare sample provides a valuable laboratory for the study of
the physics  and astronomy of magnetic reconnection events.   This
study (Paper I) examines flare morphologies, and provides general
comparison of COUP flare characteristics with those of other
active X-ray stars and the Sun. Paper II concentrates on
relationships between flare behavior, protoplanetary disks (both
passive and accreting), and other stellar properties including
rotational periods and Keplerian corotation radii. Paper II
further investigates super-hot COUP flares and magnetic field
strength on Orion T-Tauri stars.

\acknowledgements KG thanks Moisey Livshits for introducing him to
the physics of solar long decay events. We thank Vinay Kashyap,
Rachel Osten, and Paola Testa for the stimulating discussions, and
Kevin Luhman for assistance with $Spitzer$-IRAC data analysis. We
also thank the referee, Jeffery Linsky, for his time and many
useful comments that improved this work. This work was supported by 
the $Chandra$ ACIS Team (G. Garmire, PI) through the SAO grant
SV4-74018. G.M. acknowledges contribution from contract ASI-INAF
I/088/06/0. This publication makes use of data products from the
Two Micron All Sky Survey (a joint project of the University of
Massachusetts and the Infrared Processing and Analysis
Center/California Institute of Technology, funded by NASA and
NSF), and archival data obtained with the {\it Spitzer Space
Telescope} (operated by the Jet Propulsion Laboratory, California
Institute of Technology under a contract with NASA).

\appendix

\section{COMPARISON OF MASME WITH TIME-RESOLVED SPECTROSCOPY
 \label{app:TRS_analysis}}

Studies of stellar flare evolution have been conducted for some
years using low-spectral resolution time series characteristic of
CCD detectors on the ASCA, Chandra, XMM-Newton, and Suzaku
satellites. Most of these studies have used a method we call Time
Resolved Spectroscopy (TRS) where the lightcurve is divided into
intervals (e.g. flare rise, flare peak, and several flare decay
intervals) and full multiparameter fits to the spectra are made in
each interval.  These fits are typically made with a code such as
XSPEC (used here and throughout the COUP effort), Sherpa, or MIDAS
with the assumption of a one- or two-temperature plasma subject to
absorption. In cases of very high signal, non-solar elemental
abundances may also be added as free parameters. Examples of such
analyses applied to powerful PMS stellar flares include the
reheated flare of the weak-lined T Tauri star V773~Tau in the
Taurus clouds seen with ASCA \citep{Tsuboi98}, the complex flare
of the protostar YLW~16A in the Ophiuchi cloud seen with Chandra
\citep{Imanishi01}, the slow-rise flare of the weak/classical T
Tauri star LkH$\alpha$~312 in the Orion B cloud seen with Chandra
\citep{Grosso04}, the strong flare in the Herbig Ae star V892~Tau
in Taurus clouds seen with XMM-Newton \citep{Giardino04}, 18 more
brightest flare-like events from T Tauri stars in Taurus clouds
analyzed in the XEST project \citep{Franciosini07}, and the study
of 32 powerful Orion Nebula Cluster flares in Orion A based on
COUP data \citep{Favata05}.

The main difference between our MASME method and the TRS method is
that we obtain the temperature and emission measure evolutions on
rapid timescales directly from the smoothed median energy and
count rate evolutions without using a fitting procedure for
analyzed data (\S \ref{sliding_kernel_method_analysis}). The
calibration curves shown in Figure \ref{fig_kt_vs_mede} are
crucial for converting median energies to plasma temperatures.
This permits estimation of temperature and emission measure
(luminosity) at many more time points using fewer counts at each
time step than possible with TRS.  No fitting based on
least-squares or maximum likelihood statistical procedures is
made. Drawbacks to MASME include the omission of characteristic
level emission and the inability to track possible $N_H$
variations; note that these capabilities of TRS are also neglected
in many previous studies.

To compare the two methods, we have conducted a TRS analysis of
the 216 COUP flares as follows.  We examine the consistency of the
two procedures, and examine the effects of neglecting the
characteristic emission.  The peak plus decay interval of each
flare is divided into a number of segments (${\rm SegNum}$), each
hosting a roughly constant number of X-ray flare counts (${\rm
<N>_{seg}}$). For the majority of flares ($\sim 54\%$), we defined
$\ge 5$ segments with $\sim 200$ counts each. For weaker flares
($\sim 28\%$), we defined $\ga 3$ segments with 100 to 200 counts
each. For the brightest flares ($18\%$), we defined $\ga 10$
segments with $\ga 300$ counts each.

Table \ref{tbl_TRS_results} presents the results of the TRS
analysis including information on lightcurve segmentation,
``characteristic'' background spectrum, inferred flare peak
temperatures, and temperature-density slopes. Several ``typical''
flares with simple single-decay slopes are presented in Figures
\ref{fig_MASME_vs_TRS_4segments}-\ref{fig_MASME_vs_TRS_10segments}.
Results are provided for cases with (``CH'') and without
(``noCH'') accounting for the characteristic non-flare emission.
The spectrum of each segment is fitted with the
WABS~$\times$~MEKAL model \citep[compatible with previous COUP
spectroscopic fits;][]{Getman05} plus a similar model of the
characteristic component in the ``CH'' cases. The absorption $N_H$
is frozen to the value obtained from the global fits of
\citet{Getman05}, and 0.3 times solar elemental abundances are
assumed. The characteristic spectrum was obtained from the $[{\rm
t_{char1}-t_{char2}}]$ segment which is the $>20$~ks lightcurve
block with the least median value of the count rate. There is, of
course, no guarantee that the characteristic emission is the same
during the flare interval.

Figure \ref{fig_MASME_vs_TRS_general} compares several outcomes of
the MASME technique with those of TRS. Panel ($a$) shows that the
inferred TRS slopes of the trajectories in the
$\log(T)-\log(EM^{1/2})$ diagram which omit the characteristic
emission ($\zeta_{TRS,noCH}$) are systematically steeper by
$\simeq 20\%$ to $\simeq 15\%$ with increasing number of TRS
segments than those which account for characteristic emission
($\zeta_{TRS,CH}$, see open boxes and the dashed regression line
in panel $a$). This arises because the flare spectrum is generally
harder than the characteristic spectrum.

There is no strong systematic difference in slope steepness when
TRS is compared to MASME: in $\sim 45\%$ ($\sim 60\%$) of all
flares the inferred average MASME slopes ($\zeta_1$ from Table
\ref{tbl_flare_prop2}) are steeper than the TRS slope
$\zeta_{TRS,noCH}$ ($\zeta_{TRS,CH}$). The MASME-TRS slope
difference is quite high ($\sim 50-65\%$) for generally weak
flares with only few TRS segments, but decreases to $\sim 20-30\%$
for generally brighter flares with the number of TRS segments $\ga
10$ (black and grey circles and lines in panel $a$). We believe
that the major reason for the observed MASME-TRS slope difference
is not the presence or absence of the ``characteristic''
background, which contributes only a few percent to the difference
(offset between the grey and black lines), but is rather due to
the poor sampling of the temperature-density trajectory slope by
the TRS method when only few TRS segments are available in the
case of weak flares.

Panel $b$ of Figure \ref{fig_MASME_vs_TRS_general} shows that the
peak flare temperatures $T_{\rm pk,TRS,noCH}$ and $T_{\rm
pk,TRS,CH}$ obtained from the TRS method are essentially the same,
and MASME-vs.-TRS temperature differences are within $20\%$. A
20\% difference in temperature corresponds to a $< 20\%$
difference in estimated loop sizes. Large scatter in temperatures
are seen for $T \ga 200$~MK; this inability to quantify very high
plasma temperatures is expected given the 8~keV high-energy limit
and low high-energy sensitivity of Chandra spectra.

Panel $c$ of Figure \ref{fig_MASME_vs_TRS_general} examines MASME
$vs.$TRS estimates of the correction factors associated with flare
prolonged heating or triggered flare events (the factor $F(\zeta)$
presented in \S \ref{loop_modeling_analysis}).  This comparison is
restricted to 108 flares with a single average decay slope
$\zeta_1$ (i.e., flares with Flag$_1=01$ in Table
\ref{tbl_flare_prop2}).  Here again the two methods generally
agree within 20\%.

We emerge with considerable confidence in the MASME approach.
Except for temperatures $T \ga 200$~MK, basic physical flare
properties such as temperature, emission measure, and luminosity
evolution are essentially the same using the TRS and MASME
methods.  Flare model parameters such as reheating corrections and
loop sizes are also reproduced.  A reasonable estimate for the
precision of loop sizes is $< 40\%$, understanding that the
underlying model of \citet{Reale97} may not be adequate for
understanding the more complex flares. The advantage of the MASME
method is its ability to treat weaker and more rapidly variable
flares than TRS.  In the case of COUP, we are able to study $>6$
times more flares than could be studied using TRS by
\citet{Favata05}.

 \section{ERROR ANALYSIS OF MASME SPECTRAL MODELING
 \label{app:MASME_errors}}

Here we describe our estimation procedure of statistical errors on
peak flare plasma temperature in our MASME modeling. This is
important both for estimation of magnetic loop sizes and for
identification of `superhot' flares.  We use the median absolute
deviation (MAD) normalized by $0.6745/\sqrt(N)$
\begin{equation}
\Delta MedE = median | E_i - MedE | /0.6745/\sqrt(N)
\end{equation}
where $|...|$ indicates absolute value, $E_i$ is the energy of
each of the $N$ X-ray photons appearing within the sliding boxcar
kernel used to calculate peak flare median energy $MedE$. The MAD
is a well-established estimate of the uncertainty of the median
which is scaled to the standard deviation when the distribution is
Gaussian \citep{Beers90}.

Figure \ref{fig_errors}$a$  compares errors on median energy
estimated using Monte-Carlo simulations for flaring PMS stars in
the Cepheus~B region \citep[grey symbols from][]{Getman06} with
errors on COUP peak flare median energy (black symbols) obtained
with the simple MAD formula above. There is a perfect match
between the two completely different methods of error analysis,
even for different $MedE$ strata (circles vs. $\times$).

\begin{figure}
\centering
\includegraphics[angle=0.,width=6.5in]{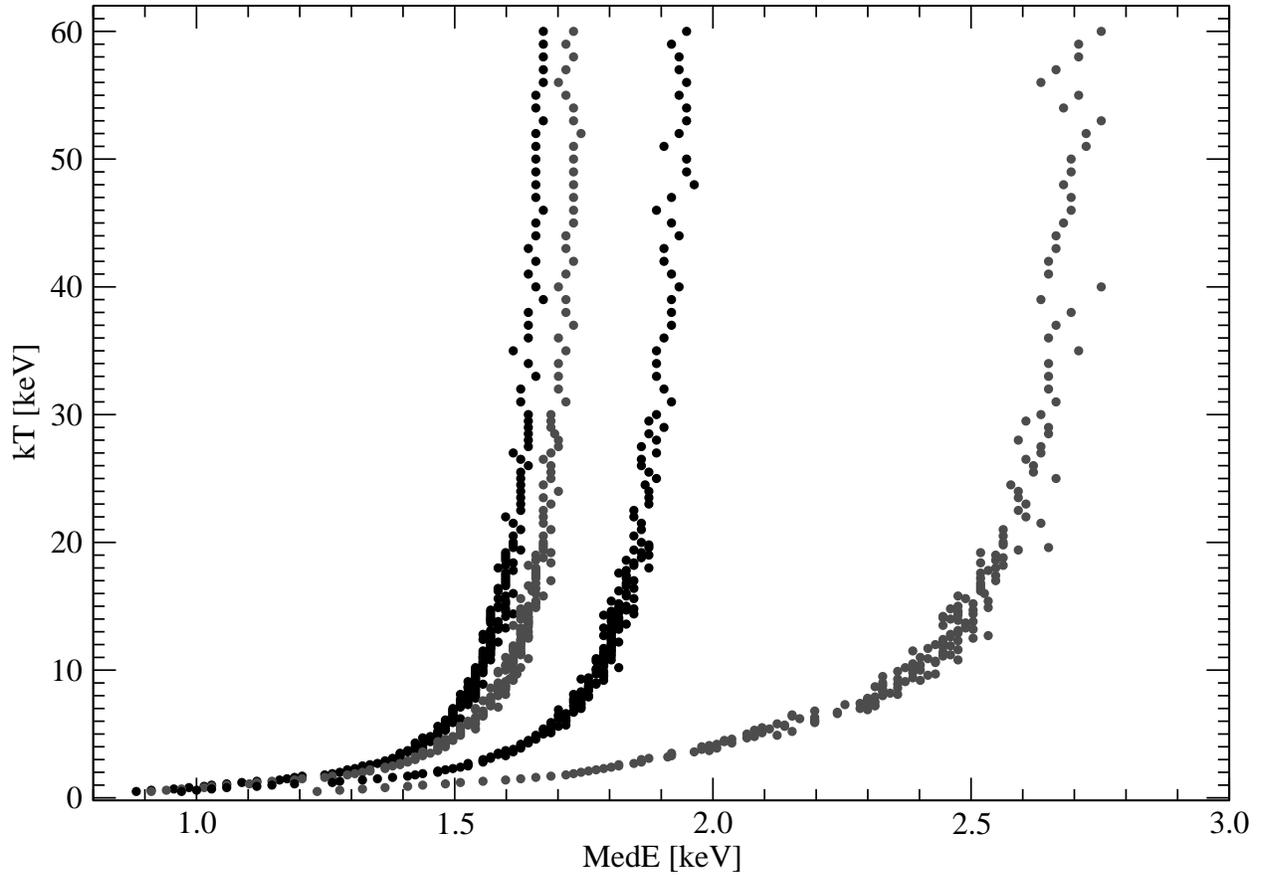}
\caption{Calibration of plasma temperature vs. observed X-ray
median energy for representative column densities ($\log
N_H=(20.5,21,21.5,22)$~cm$^{-2}$, from left to right) calculated
from spectral simulations. \label{fig_kt_vs_mede}}
\end{figure}

\begin{figure}
\centering
\includegraphics[angle=0.,width=6.5in]{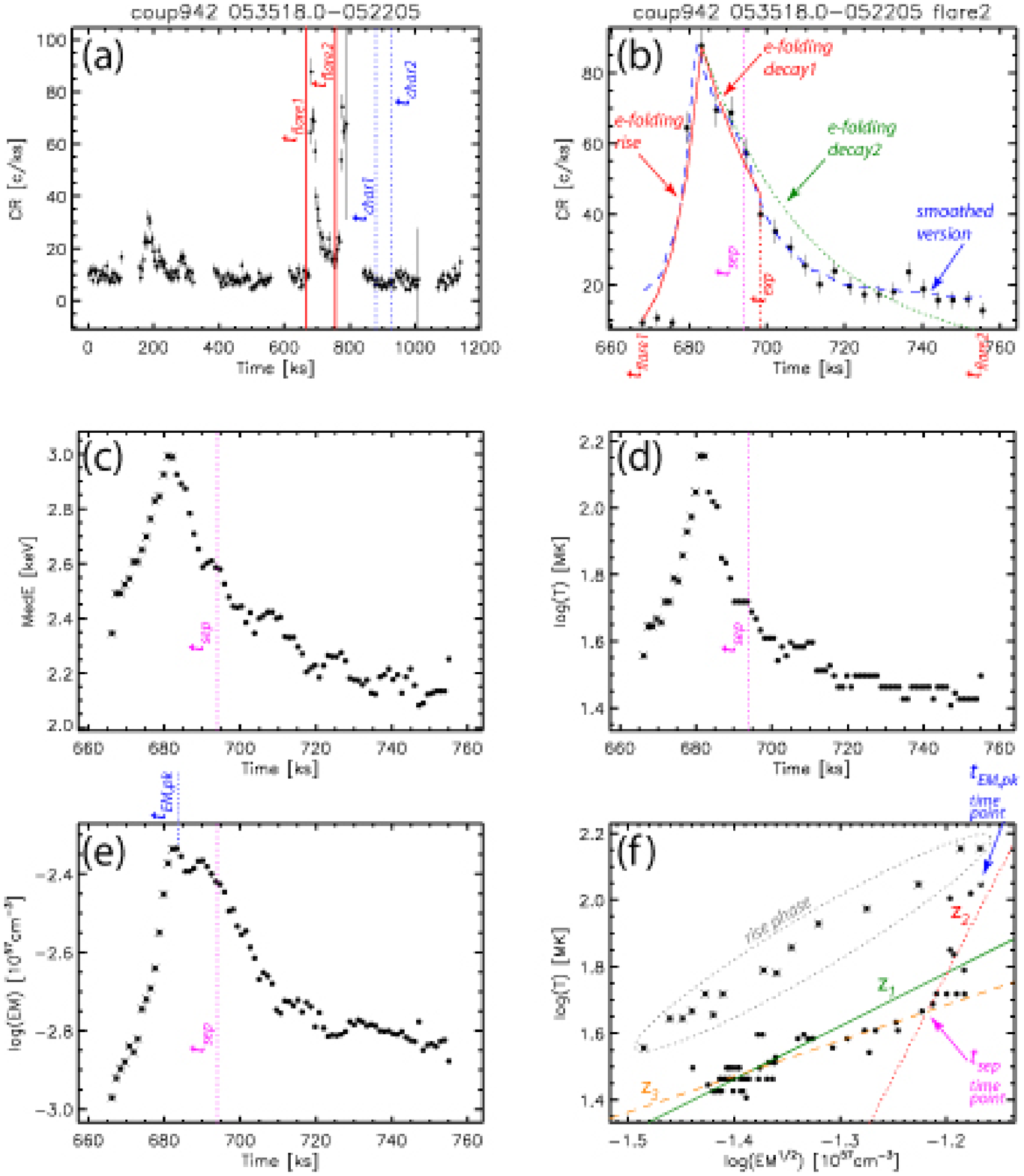}
\caption{Illustration of our X-ray spectral modeling procedure
using flare \#2 of COUP \#942 using the method of adaptively
smoothed median energy (MASME). Details on specific symbols and
lines are provided in \S \ref{derived_quantities_analysis}. (a)
X-ray lightcurve over the whole COUP observation (six nearly
consecutive exposures) from the COUP Atlas of \citet{Getman05}.
(b) Lightcurve of the flare in binned histogram (black) and
adaptively smoothed (blue) representations. (c) Evolution of the
median energy. (d) Evolution of the plasma temperature. (e)
Evolution of the emission measure. (f) Evolution in the $\log T -
\log \sqrt(EM)$ plane. [{\it See the electronic edition of the journal for the full COUP flare atlas, Figs. 2.2-2.216}]. \label{fig_flare_example}}
\end{figure}

\begin{figure}
\centering
\includegraphics[angle=0.,width=6.5in]{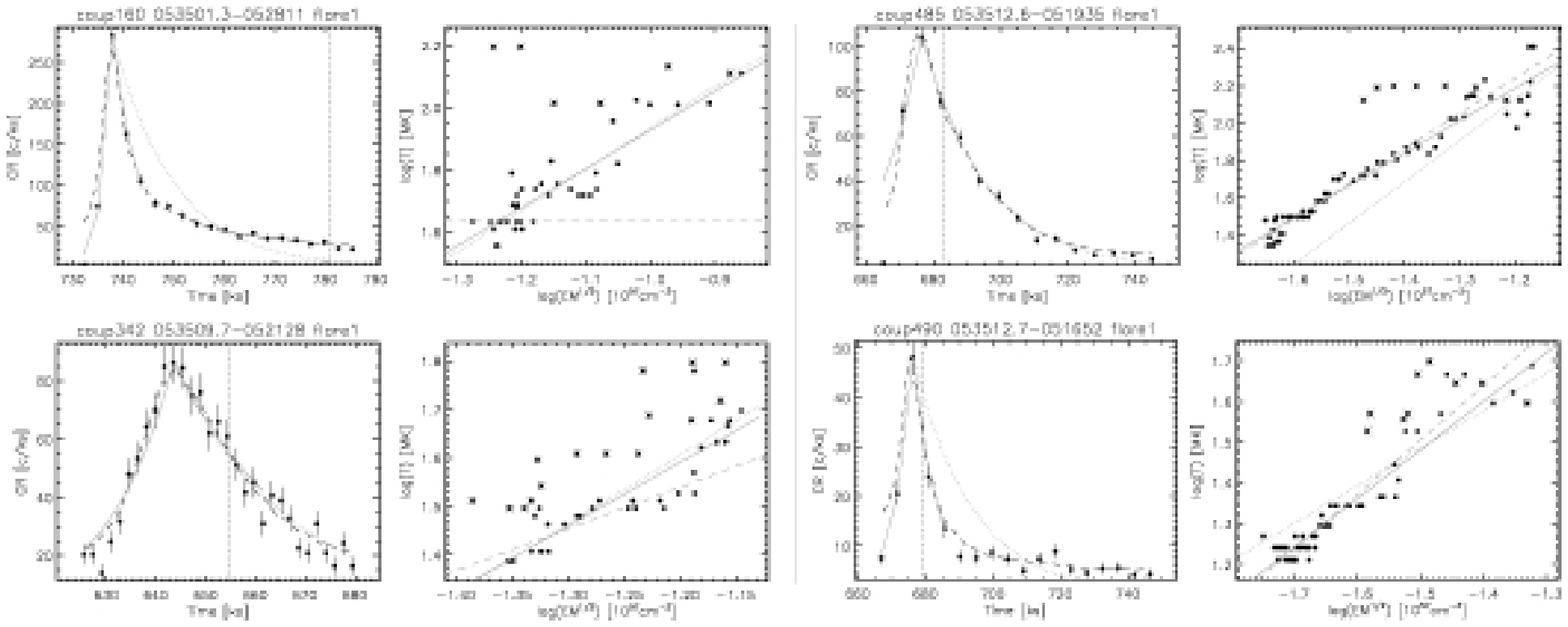}
\caption{Lightcurves and inferred temperature-density trajectories
for four representative flares in the class ``typical''.
\label{fig_morphology_typical}}
\end{figure}

\begin{figure}
\centering
\includegraphics[angle=0.,width=6.5in]{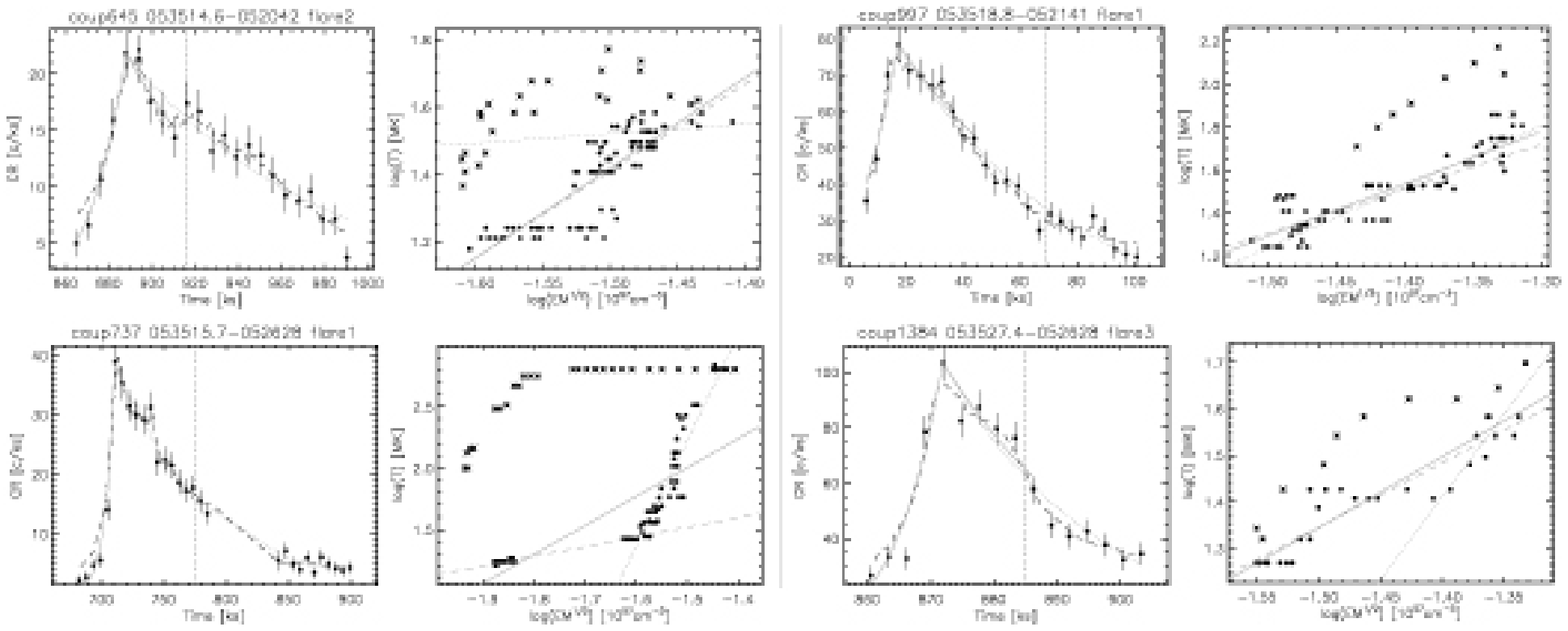}
\caption{Lightcurves and inferred temperature-density trajectories
for four representative flares in the class ``step''.
\label{fig_morphology_step}}
\end{figure}

\begin{figure}
\centering
\includegraphics[angle=0.,width=6.5in]{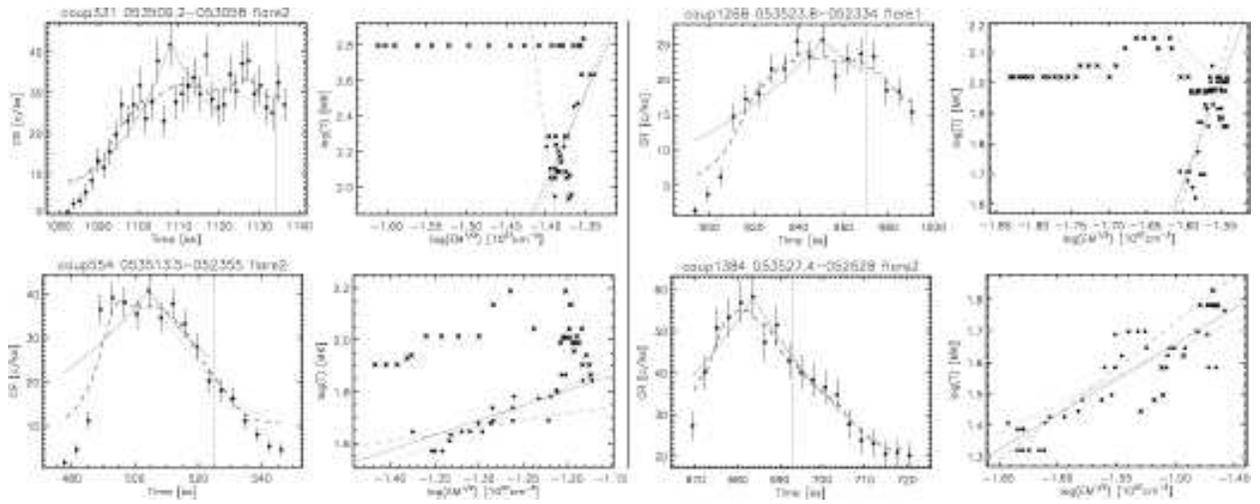}
\caption{Lightcurves and inferred temperature-density trajectories
for four representative flares in the class
``slow-rise-and/or-top-flat''. \label{fig_morphology_srf}}
\end{figure}

\begin{figure}
\centering
\includegraphics[angle=0.,width=6.5in]{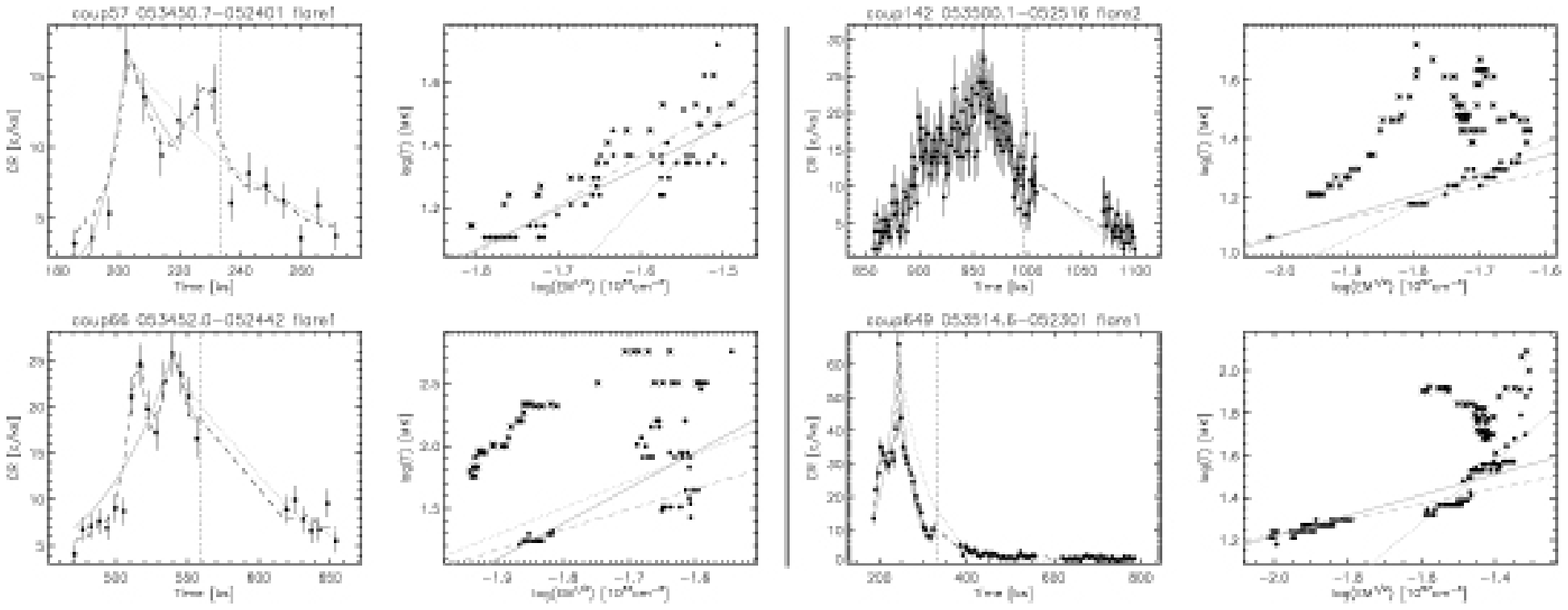}
\caption{Lightcurves and inferred temperature-density trajectories
for four representative flares in the class ``double''.
\label{fig_morphology_dbl}}
\end{figure}

\begin{figure}
\centering
\includegraphics[angle=0.,width=7.5in]{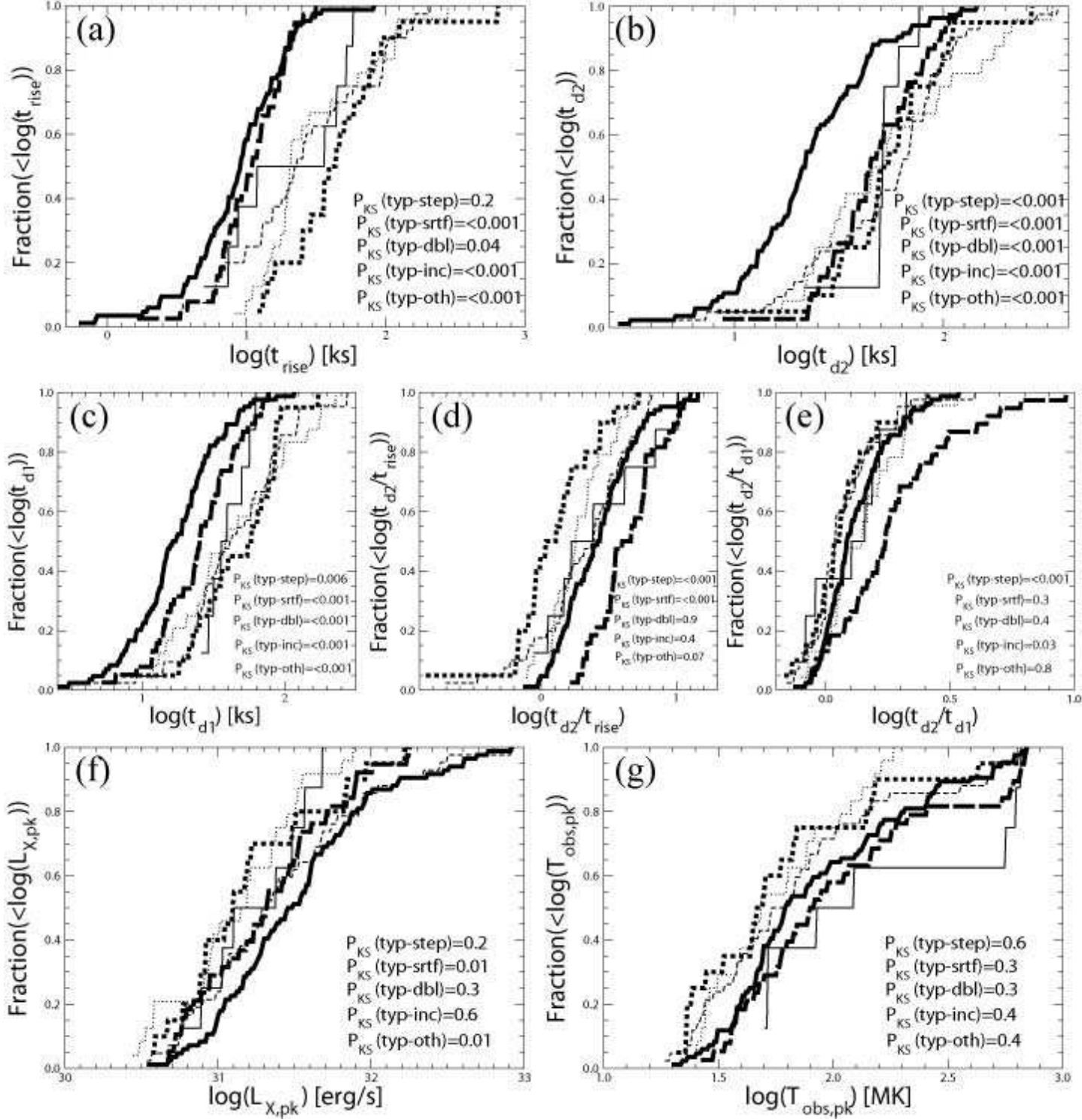}
\caption{\footnotesize Cumulative distributions of flare
properties: ($a$) rise time $\tau_{rise}$, ($b$) decay time
$\tau_{d2}$, ($c$) decay time $\tau_{d1}$, ($d$) ratio
$\tau_{d2}/\tau_{rise}$, ($e$) ratio $\tau_{d2}/\tau_{d1}$, ($f$)
peak luminosity, and ($g$) peak temperature.  Line types indicate
morphological classes: 84 ``typical'' flares (thick solid line),
38 ``step'' (thick dashed), 20 ``srtf'' (thick dotted), 8
``double'' (thin solid), 42 ``incomplete'' (thin dashed), and 24
``other'' (thin dotted). K-S probabilities test the assumption
that ``typical'' and other flare types are drawn from the same
underlying distribution. \label{fig_global_props}}
\end{figure}

\begin{figure}
\centering
\includegraphics[angle=0.,width=7.5in]{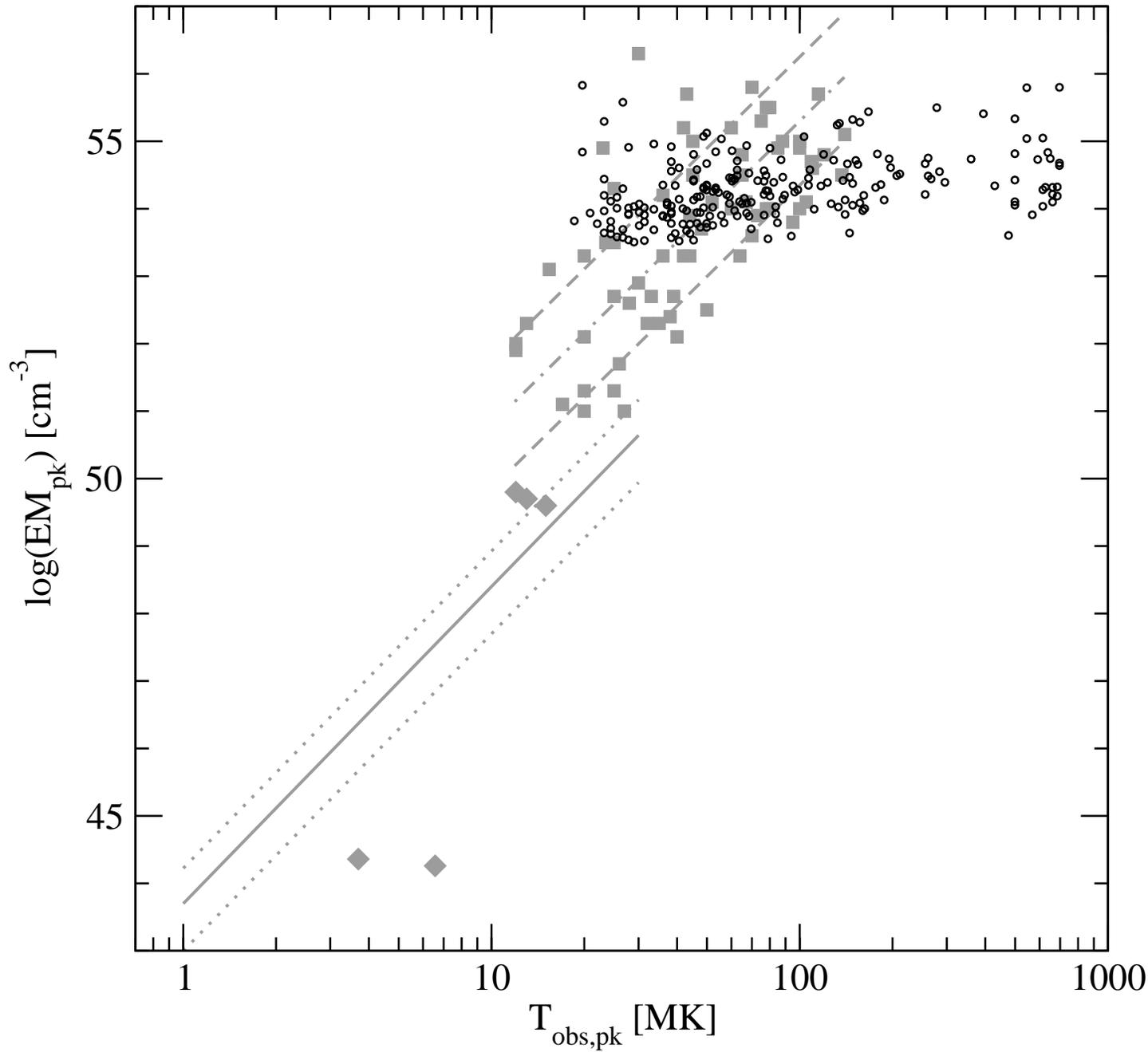}
\caption{\footnotesize Plot of flare peak emission measure $vs.$
peak temperature for different flare classes:  216 COUP flares
(black circles); stellar flares compiled by \citet{Gudel04} (grey
boxes) with their linear regression fit and 1$\sigma$-ranges
(dashed-dotted and dashed lines); and the regression fit to solar
flares compiled by \citet{Aschwanden08} (solid and dotted lines).
Representative solar LDEs discussed in the text are shown as grey
diamonds. \label{fig_compare_older_stars1}}
\end{figure}

\begin{figure}
\centering
\includegraphics[angle=0.,width=7.5in]{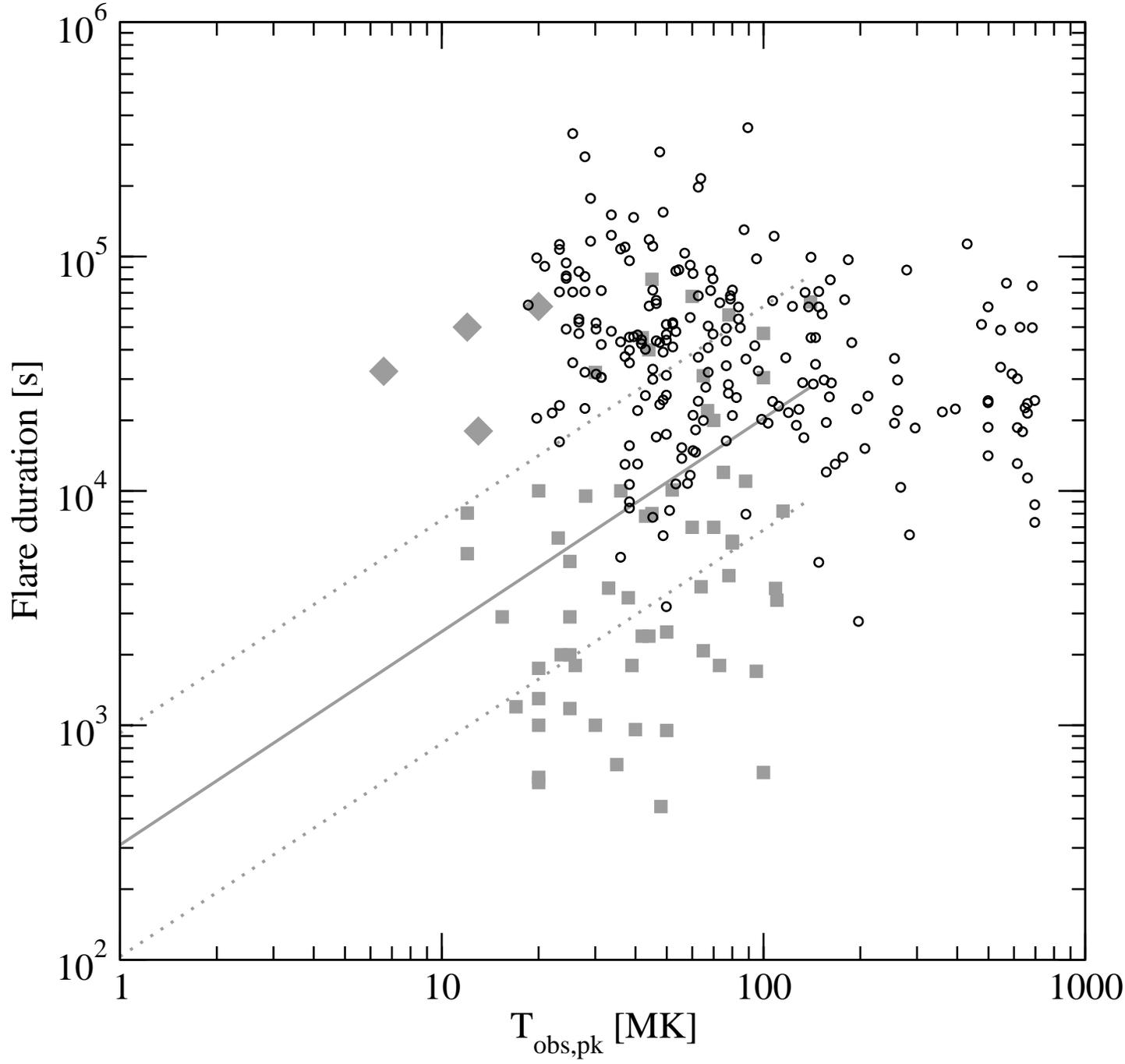}
\caption{\footnotesize  Plot of flare duration $vs.$ flare peak
temperature for different flare classes.  Symbols and lines follow
those of Figure~\ref{fig_compare_older_stars1}.
\label{fig_compare_older_stars2}}
\end{figure}

\begin{figure}
\centering
\includegraphics[angle=0.,width=7.5in]{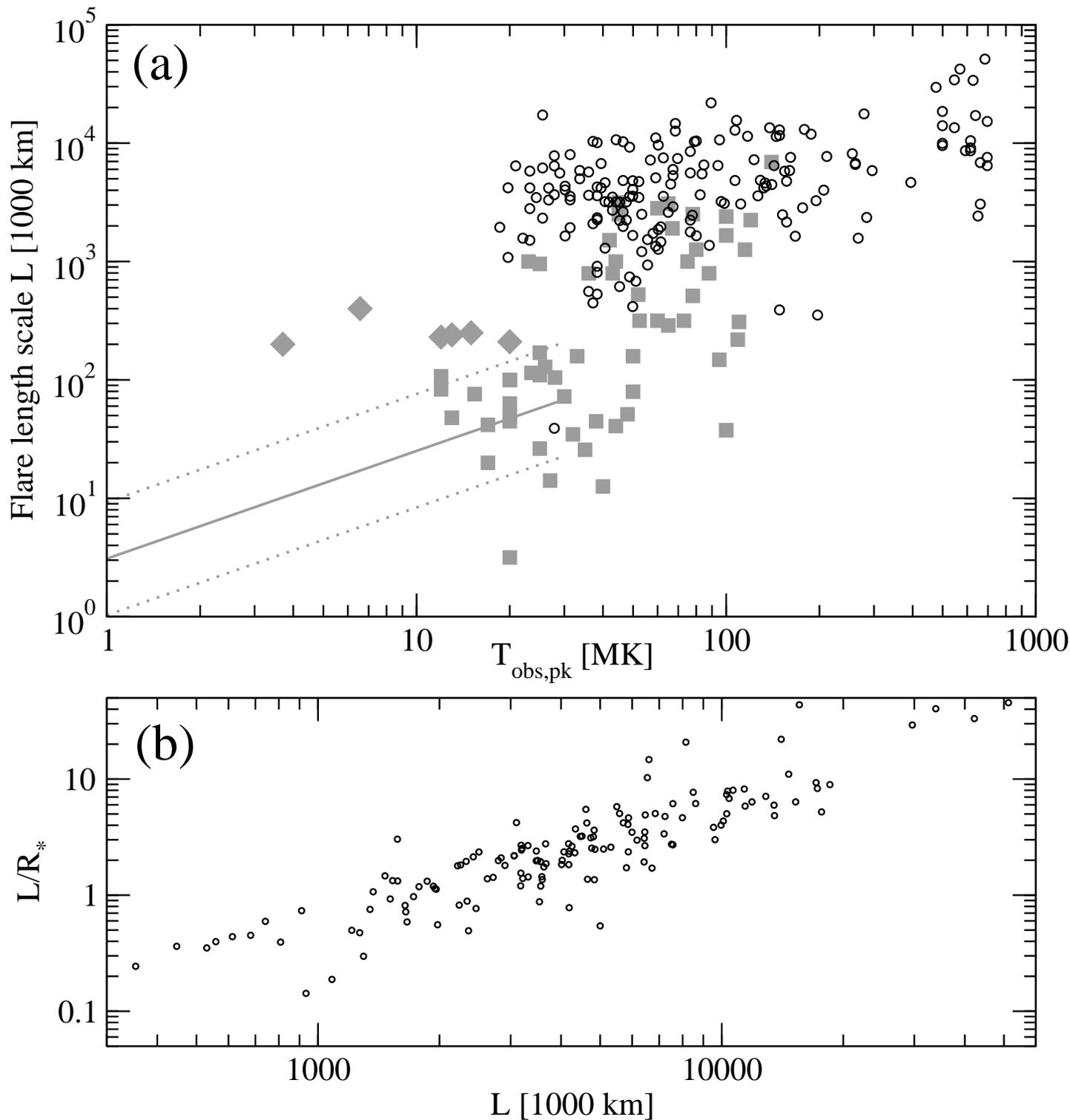}
\caption{\footnotesize ($a$) Flare length scale $vs.$ flare peak
temperature for different flare classes. Symbols and lines follow
those of Figure~\ref{fig_compare_older_stars1}. ($b$) Comparison
of inferred COUP flaring loop length to the loop length normalized
by stellar radius ($L/R_{\star}$) for 147 COUP stars with known
radii. \label{fig_compare_older_stars3}}
\end{figure}

\begin{figure}
\centering
\includegraphics[angle=0.,width=6.0in]{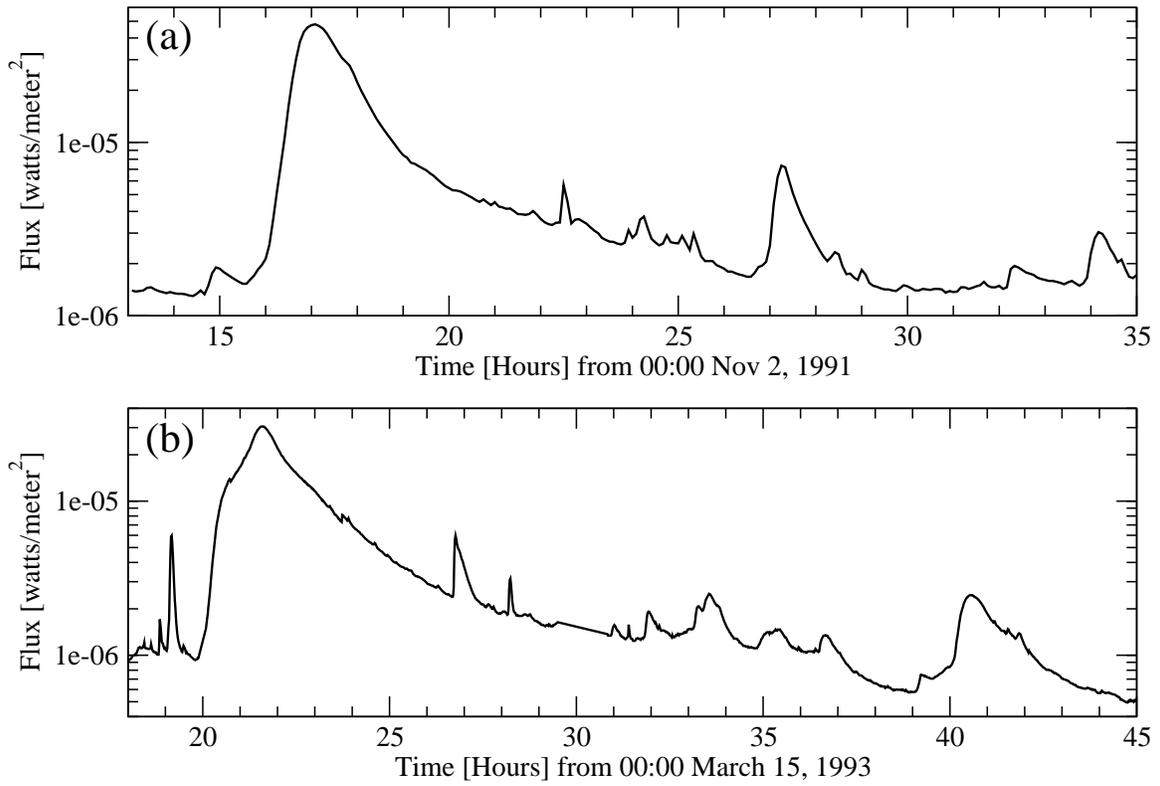}
\caption{\footnotesize X-ray lightcurves obtained in $1-8$~\AA
band by GOES satellites during formation of giant solar X-ray
arches \citep{Svestka95} on:  ($a$) 2 Nov 1991, and ($b$) 15 Mar
1993. \label{fig_goes_lde}}
\end{figure}

\begin{figure}
\centering
\includegraphics[angle=0.,width=6.5in]{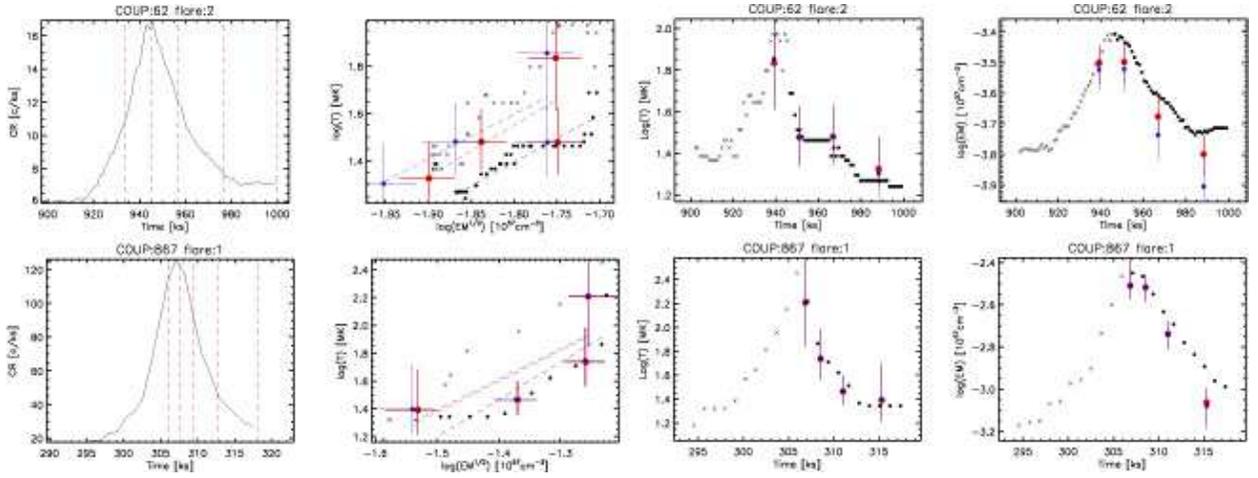}
\caption{\footnotesize Comparison of MASME and TRS flare analyses
for representative ``typical''  flares with $4$ TRS segments. Each
source is represented by four panels (from left to right):
adaptively smoothed lightcurve with dashed red lines indicating
TRS segments; $\log(T)-\log(EM^{1/2})$ diagram; temporal evolution
of plasma temperature;  and temporal evolution of emission
measure. Black circles ($\times$s) indicate MASME-derived decay
(rise) points.  Blue (red) circles with error bars are TRS-derived
points for the peak-decay lightcurve phase with (without)
accounting for the ``characteristic'' background. Dashed lines are
linear regressions to MASME points over the whole decay phase
(black; denoted as slope $\zeta_1$ in Table \ref{tbl_flare_prop2})
and to TRS points over the peak-decay phase (blue and red for the
``CH'' and ``noCH'' cases, respectively).
\label{fig_MASME_vs_TRS_4segments}}
\end{figure}

\begin{figure}
\centering
\includegraphics[angle=0.,width=7.0in]{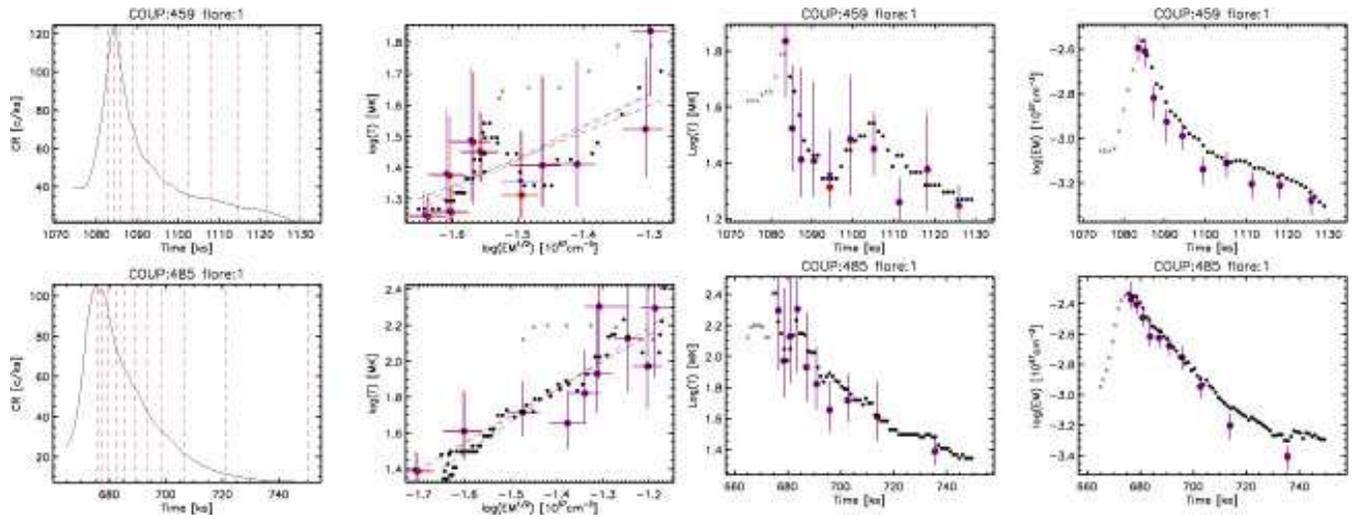}
\caption{Comparison of MASME and TRS flare analyses for
representative ``typical'' flares with $10$ TRS segments.  See
Fig. \ref{fig_MASME_vs_TRS_4segments} for details.
\label{fig_MASME_vs_TRS_10segments}}
\end{figure}

\begin{figure}
\centering
\includegraphics[angle=0.,width=6.5in]{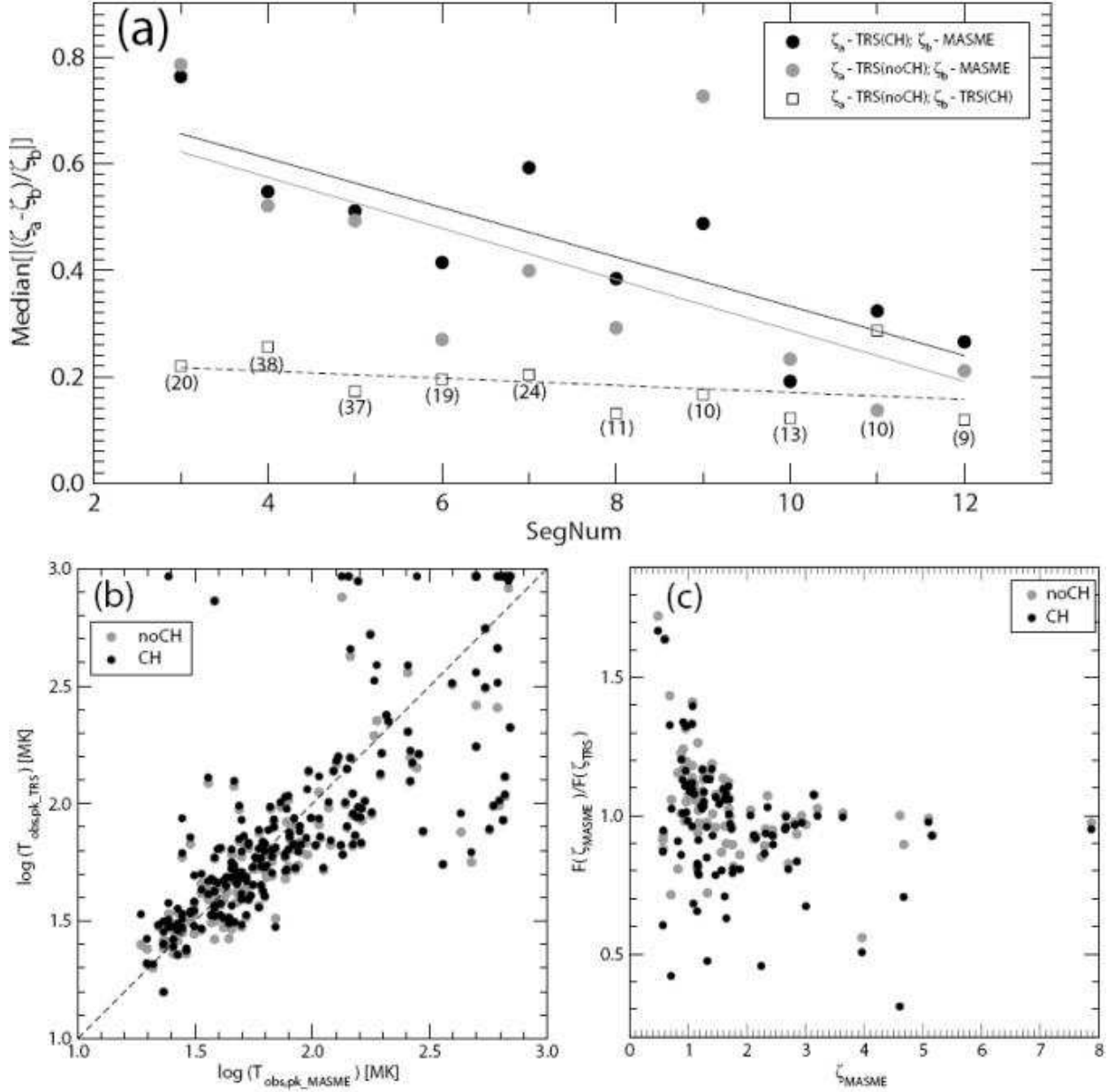}
\caption{\footnotesize Comparison of MASME and TRS flare analyses
results. ($a$) Median fractional difference between inferred
slopes in the $\log(T)-\log(EM^{1/2})$ diagram as a function of
number of TRS segments. Black circles indicate correction for
``characteristic'' background emission (``CH'') while grey circles
indicate no correction (``noCH'').  Open boxes indicate median
fractional difference between TRS ``CH'' and TRS ``noCH''. Lines
represent linear regressions of the data points. Numbers in
parentheses report the number of flares that have the specified
number of TRS segments. ($b$) Comparison between MASME- and
TRS-inferred peak flare temperatures. ($c$) Ratio of the MASME to
TRS correction factors $F(\zeta)$ for prolonged heating (see \S
\ref{loop_modeling_analysis}) versus the MASME slope.
 \label{fig_MASME_vs_TRS_general}}
\end{figure}

\begin{figure}
\centering
\includegraphics[angle=0.,width=6.0in]{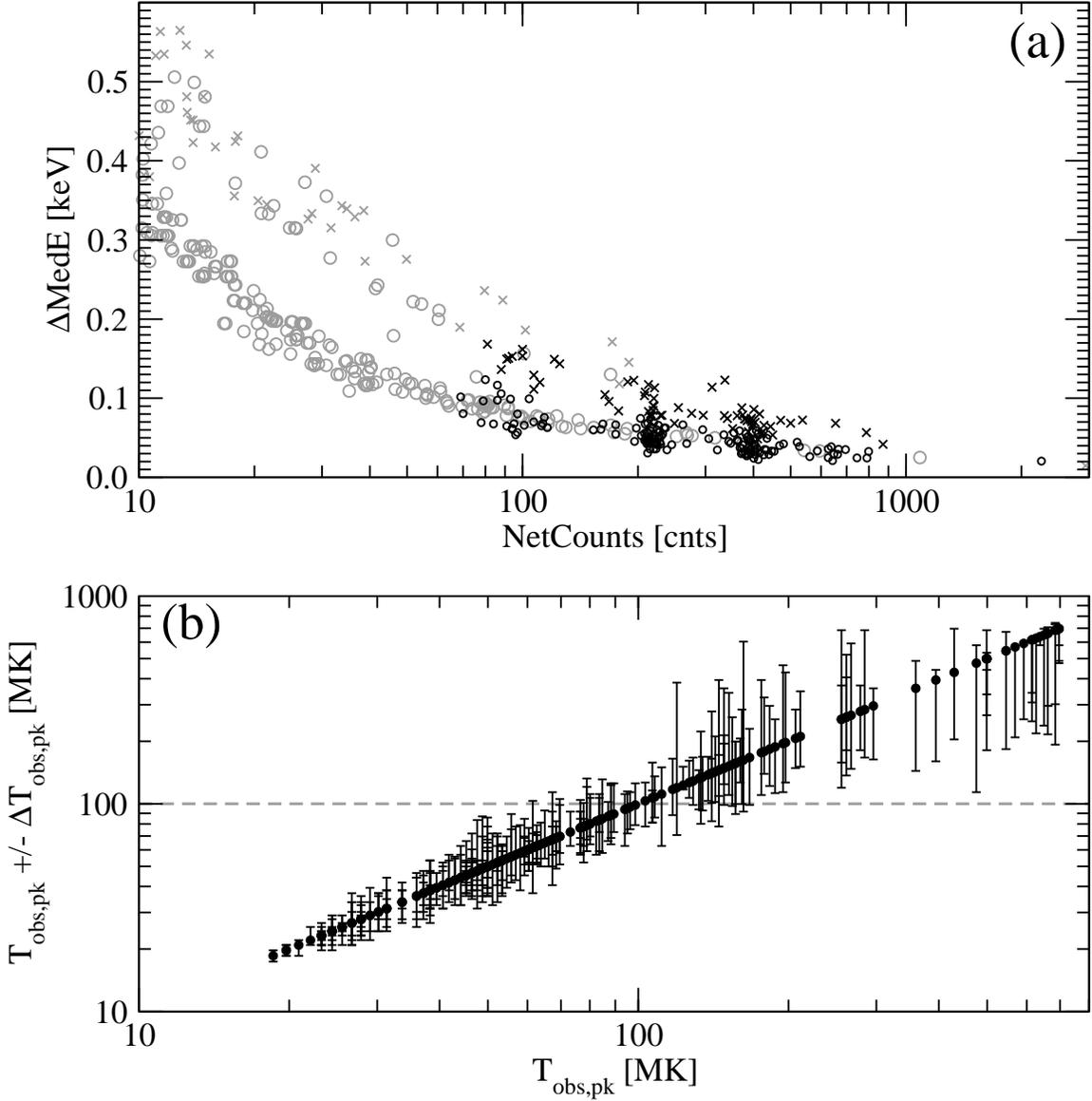}
\caption{Validation of median energy and peak temperature
estimates. ($a$) Comparison between 1$\sigma$ errors on median
energy obtained from Monte-Carlo simulations of Cepheus~B X-ray
sources \citep[grey symbols from][]{Getman06} and errors derived
from the MAD technique applied to COUP flaring sources (black
symbols). Circles indicate $MedE<2$~keV and crosses indicate $MedE
>2$~keV. ($b$) Errors on peak flare plasma temperature derived for
our COUP flare sample based on the median energy calibration curve
in Figure \ref{fig_kt_vs_mede}. Upper errors on $T_{obs,pk} \ga
300$~MK are unreliable due to the lack of calibration data points
above 700~MK . \label{fig_errors}}
\end{figure}

The errors on peak flare plasma temperature are shown in Figure
\ref{fig_errors}$b$. They are derived through propagation of
errors on peak flare $MedE$ using the simulation-based
$T_{obs,pk}-MedE$ calibration curves (Figure
\ref{fig_kt_vs_mede}). Upper errors for many sources with
$T_{obs,pk} \ga 300$~MK are unreliable due to the lack of
calibration data above 700~MK (Figure \ref{fig_kt_vs_mede}). Lower
1$\sigma$ errors are generally $<20\%$ for $T_{obs,pk} < 50$~MK,
$<30\%$ for $T_{obs,pk} < 100$~MK, $<40\%$ for $T_{obs,pk} <
200$~MK, and $\sim 60\%$ for the hottest flares.

\begin{figure}
\centering
\includegraphics[angle=0.,width=7.2in]{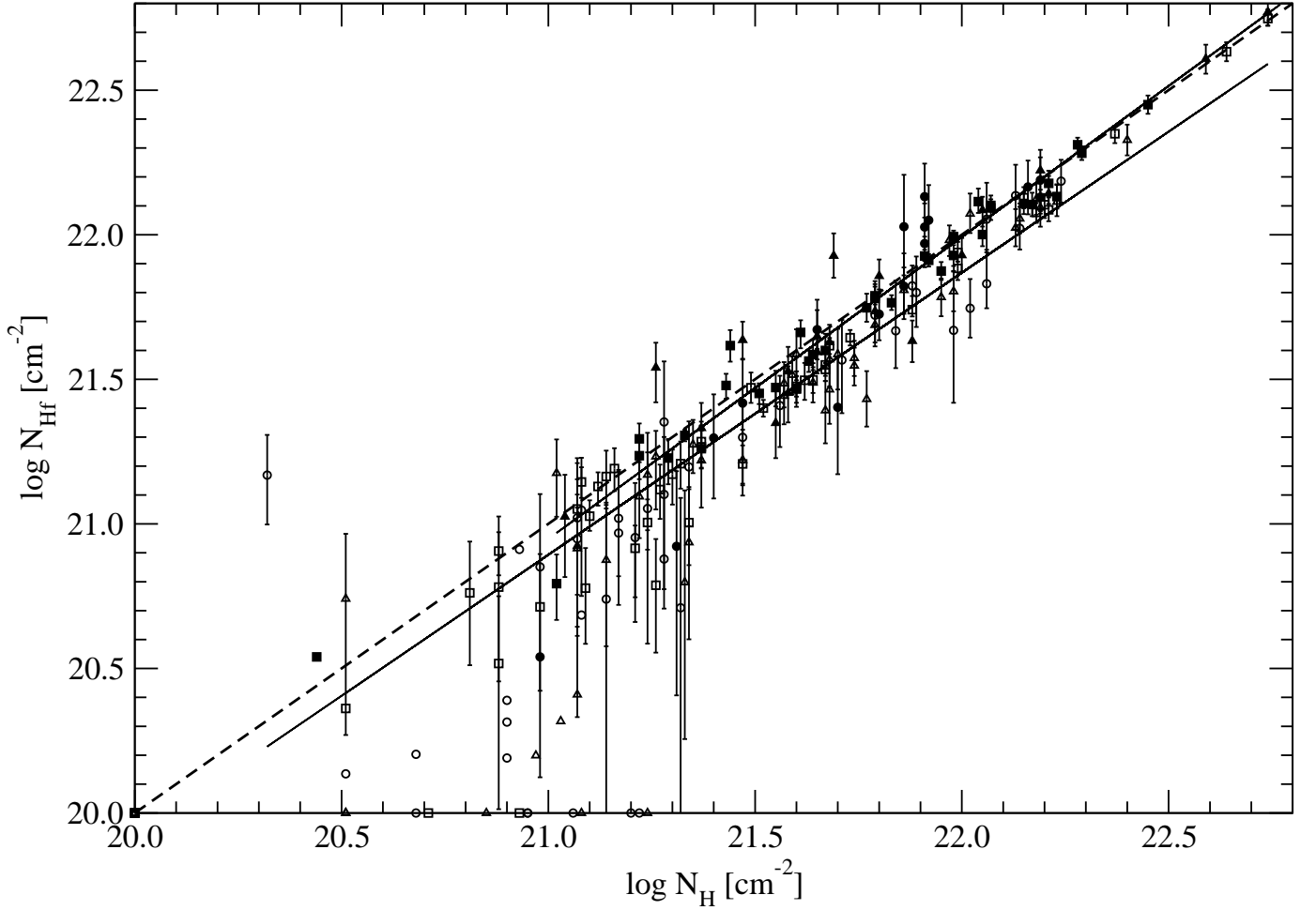}
\caption{Comparison of column densities derived here from fits of
flare spectra, $N_{Hf}$, with column densities  \citep{Getman05} derived
from fits of time-integrated COUP spectra (within the whole COUP observation),
$N_{H}$. Symbols represent different flare counts:
circles ($NCf<1000$), triangles ($1000<NCf<2000$)
and squares ($NCf>2000$). Filled symbols denote 73 super-hot flares.
The dashed line shows values of $\log(N_{Hf})-\log(N_{H})=0$. The shorter
solid line is a regression line to all super-hot flares with known formal
statistical errors on $N_{Hf}$ (vertical bars). The longer solid line is a
regression line to all cooler flares with known errors.
\label{fig_syst_errors}}
\end{figure}

We now further consider how uncertainties of inferred X-ray column
densities may affect the errors of derived flare peak plasma
temperatures shown in Figure 15b. First, it is important to note
that 89\% of super-hot flares (compared to 50\% for cooler flares)
are found in stars for which their inferred X-ray column densities
are $\log N_H > 21.3$~cm$^{-2}$ (Figure \ref{fig_syst_errors}). 
This indicates that the super-hot
flares are not a result of some data analysis bias, such as an
effect of a systematic underestimate of $N_H$, but rather related
to real physical phenomena relating to actively accreting disks
around young stars, as it is shown in Paper~II.

Second, because the studied COUP stars are extremely strong
sources with thousands of counts, statistically, their $\log N_H$
is measured with very high accuracy.  Formal statistical
uncertainties are less than $\pm 0.03$ dex (1-sigma) for sources
with $\log N_H > 21.2$~cm$^{-2}$ and less than $\pm 0.07$ dex for
most of the softer COUP sources. Such statistical errors on $N_H$
have only marginal effects on the resulting temperature errors.
For example, in a representative case of a super-hot flare from
the COUP source \#1309 with the $\log N_H = 22.00 \pm
0.02$~cm$^{-2}$, the propagation of statistical errors on $N_H$
will change the errors on plasma temperature of
$T_{obs,pk}=188_{-67}^{+75}$~MK shown in Figure 15b from 40\% to
53\% and from 36\% to 43\% of an inferred mean temperature value
for the upper and lower error limits, respectively. For a
representative case of a cooler flare from COUP source \#1492 with
$\log N_H = 21.07 \pm 0.05$~cm$^{-2}$, its errors on temperature
of $T_{obs,pk}=51_{-19}^{+21}$~MK will change from 41\% to 50\%
and from 37\% to 38\%, for the upper and lower error limits,
respectively.

However, for the group of 8 extremely soft sources\footnote{These
are COUP stars \#\# 71, 152, 394, 597, 971, 1481, 1516, 1595.}
with their reported $\log N_H$ values truncated at
$20.0$~cm$^{-2}$, the spectral model assumed by Getman et al.
(2005) used to derive $\log N_H$ (two plasma MEKAL temperatures
with 0.3$\times$solar elemental abundances with Wisconsin gas
absorption) is not sufficiently complex to fit the observed
spectra well around the O~VII and O~VIII emission line complex.
The reason is that spectral lines are present from FIP-related
abundance anomalies which are not in the Getman et al. spectral
model. These abundance effects are documented in detail by
\cite{Maggio07} for nearly the same sample of bright COUP stars as
we analyze here. We re-fitted these 8 extremely soft spectra with
two-temperature component VMEKAL models with individual elemental
abundances as free parameters to obtain statistically good
spectral fits. In accord with the results of Maggio et al. who
used VAPEC models, we find that while for some of these spectra
(\#\# 71, 597, 1481) their formal $\log N_H$ reported by XSPEC
continue to be less than 20.0, for others their inferred $\log
N_H$ may increase to 20.5-20.7~cm$^{-2}$. In the former cases, our
calibration curves in Figure~1 assuming $\log N_H =
20.0$~cm$^{-2}$ will not change due to the insensitivity of ACIS-I
spectra to very low column densities.  In the latter cases, the
shift from $\log N_H =20.0$ to $20.5-20.7$~cm$^{-2}$ only slightly
affects the lower error limits on peak flare plasma temperatures,
which are typically around $30$~MK. For example, the lower error
on the plasma temperature $T_{obs,pk} = 27_{-6}^{+7}$~MK of the
source \# 1595 will change only from 22\% to 26\% of an inferred
mean temperature value when the shift from $\log N_H =
20$~cm$^{-2}$ towards newly inferred $20.5$~cm$^{-2}$ is applied.

Time-integrated COUP spectra are typically several times stronger than
spectra of individual flares. To take advantage of this fact in our
flare analysis, we fixed column densities to $N_H$ values derived
from the two-temperature fits of time-integrated COUP spectra \citep{Getman05}.
To verify if this produces a source of systematic uncertainty on
column density, flare spectra were extracted within flare time range of
$[\rm{t}_{flare1}-\rm{t}_{flare2}]$ and fitted by one-temperature WABS$\times$MEKAL
model with 0.3$\times$solar abundances allowing both temperature and
column density to be free parameters. Figure \ref{fig_syst_errors}
comparing column densities resulted from these flare fits, $N_{Hf}$,
with column densities from time-integrated fits, $N_H$, shows no systematic
differences in the case of super-hot flares (shorter solid regression line vs.
dashed line), but suggests that $N_H$ can be systematically overestimated
by $\sim 0.12$~dex in the case of ``cooler'' flares (longer solid regression
line vs. dashed line). This has only marginal effect on resulting
temperature errors. For example, in a representative case of a ``cooler''
flare from COUP source \#1492 considered above, this systematic shift
will only change the temperature upper limit from 50\% to 55\% of an
inferred mean temperature value of $51$~MK.

Finally, systematic uncertainty on column densities may arise from
performance uncertainties of the $Chandra$-ACIS detector. Such uncertainties
have been evaluated by \citet{Drake06}. They show that in $10^4$ source
count regime appropriate for COUP flares, instrumental uncertainties for unabsorbed
($\log N_H \sim 20$~cm$^{-2}$) sources may become comparable to (but not exceed)
statistical uncertainties. For example, their spectral simulations predict a $0.07$~dex
instrumental systematic error on $N_H$, which is similar to our statistical 
errors of $\la 0.07$~dex for soft COUP sources.

We thus find that the typical flare peak temperature errors shown
in Figure~15b (typically 20-40\%) could be slightly larger
(typically 25-55\%) if statistical and systematic uncertainties of
$\log N_H$ values are applied.  There is no evidence that the
super-hot temperatures are an artifact of our analysis procedures
or uncertainties. From equations (1)$-$(3), 25-55\% errors on peak flare temperature
result in 15-30\% errors on inferred loop sizes.

As with median energies, we expect the statistical errors on COUP
peak flare X-ray luminosities to be comparable with those obtained
from Monte-Carlo simulations of Cepheus~B sources which are $\la
0.2$ dex \citep[see bottom-right panel in Figure~12
in][]{Getman06}.

\clearpage

\begin{deluxetable}{cccccccccccccccc}
\centering \rotate \tabletypesize{\tiny} \tablewidth{0pt}
\tablecolumns{16}

\tablecaption{Flare properties 1 \label{tbl_flare_prop1}}
\tablehead{

\colhead{Source} & \colhead{FlareNo} & \colhead{NCf} &
\colhead{Bgf} & \colhead{MEf} & \colhead{t$_{char1}$} &
\colhead{t$_{char2}$} & \colhead{t$_{flare1}$} &
\colhead{t$_{flare2}$} & \colhead{CR$_{char}$} &
\colhead{CR$_{flare,pk}$} & \colhead{t$_{CR,pk}$} &
\colhead{t$_{T,pk}$} & \colhead{t$_{EM,pk}$} & \colhead{t$_{exp}$}
& \colhead{t$_
{sep}$}\\

&&(cnts)&(cnts)&(keV)&(ks)&(ks)&(ks)&(ks)&(cnts/ks)&(cnts/ks)&(ks)&(ks)&(ks)&(ks)&(ks)\\

(1)&(2)&(3)&(4)&(5)&(6)&(7)&(8)&(9)&(10)&(11)&(12)&(13)&(14)&(15)&(16)}

\startdata
       7 &        1 &  3209 &  11.7 & 1.31 &   60.0 &  100.0 &  930.0 &  967.0 & 38.3 & $ 164.0\pm11.5$ &  951.1
&  948.9 &  951.2 &  967.0 &  960.0\\
       9 &        1 &  3980 &   6.1 & 1.55 &  160.0 &  200.0 &  523.0 &  558.0 & 12.8 & $ 210.0\pm 7.7$ &  542.4 &
538.5 &  544.2 &  553.0 &  549.0\\
      11 &        1 &  1158 &   3.5 & 2.17 &  883.0 &  940.0 &  977.0 & 1009.0 &  3.4 & $ 105.5\pm 6.3$ &  986.9
&  978.6 &  986.5 &  993.0 &  997.5\\
      23 &        1 &  4370 &  13.5 & 1.37 &  440.0 &  480.0 &  740.0 &  788.0 & 51.8 & $ 306.7\pm15.5$ &  747.0
&  746.0 &  747.1 &  752.0 &  750.0\\
      27 &        1 &   350 &   5.9 & 1.37 &  220.0 &  260.0 &  258.0 &  302.0 &  4.4 & $  14.2\pm 1.8$ &  282.6 &
283.2 &  285.4 &  295.0 &  295.0\\
      27 &        2 &  1653 &  28.1 & 1.31 &  220.0 &  260.0 &  681.0 &  890.0 &  4.4 & $  19.5\pm 2.0$ &  710.8 &
720.9 &  711.8 &  743.0 &  743.0\\
      28 &        1 & 13308 &  22.9 & 1.44 &  200.0 &  300.0 &  615.0 &  790.0 &  3.6 & $ 300.3\pm10.7$ &  623.6
&  616.0 &  622.9 &  680.0 &  745.0\\
      43 &        1 &  2744 &   7.5 & 1.42 &  470.0 &  520.0 &  665.0 &  800.0 &  4.3 & $  39.4\pm 2.8$ &  716.5 &
709.5 &  714.1 &  770.0 &  730.0\\
\enddata

\tablecomments{Column 1: COUP source number from \citet{Getman05}.
Column 2: Flare number. Columns 3-5: Estimated net flare counts,
background counts during the flare, and median X-ray energy of
flare counts. Columns 6-7: Start and stop times for measurement of
the ``characteristic'' state.  All times  are measured from the
start time of the COUP observation. Columns 8-9: Observed start
and stop times of the flare. Column 10: Median count rate of the
``characteristic'' state. Column 11: Observed flare peak count
rate. Columns 12-14: Time points when flare count rate, plasma
temperature, and emission measure have maximum values. Column 15:
Time point until which the lightcurve exhibits a pure exponential
decay. Column 16: Separation time point between two distinct decay
phases. This is relevant for cases with double slope decays (see
Flag$_1$ from Table \ref{tbl_flare_prop2}).}

\end{deluxetable}

\clearpage \clearpage

\begin{deluxetable}{ccccccccccccccccc}
\centering \rotate \tabletypesize{\tiny} \tablewidth{0pt}
\tablecolumns{17} \tablecaption{Flare properties 2
\label{tbl_flare_prop2}} \tablehead{

\colhead{Source} & \colhead{FlareNo} & \colhead{$\tau_{r}$} &
\colhead{$\tau_{d1}$} & \colhead{$\tau_{d2}$} & \colhead{$\log
L_{X,pk}$} & \colhead{T$_{obs,pk}$} & \colhead{EM$_{pk}$} &
\colhead{T$_{EM,pk}$} & \colhead{$\zeta_{1}$} &
\colhead{$\zeta_{2}$} & \colhead{$\zeta_{3}$} & \colhead{L$_{1}$}
& \colhead{L$_{2}$} & \colhead{L$_{lim}$} & \colhead{Flag$_{1}
$} & \colhead{Type}\\

&&(ks)&(ks)&(ks)&(erg~s$^{-1}$)&(MK)&($10^{53}$~cm$^{-3}$)&(MK)&&&&($10^{10}$~cm)&($10^{10}$~cm)&
($10^{10}$~cm)&&\\

(1)&(2)&(3)&(4)&(5)&(6)&(7)&(8)&(9)&(10)&(11)&(12)&(13)&(14)&(15)&(16)&(17)}

\startdata
 7 & 1 & 13.5 & 13.0 & 13.0 & 31.62 & 40.62 & 40.37 & 27.85 & 0.6 & 1.3 & -1.1 & 12.97 & 12.97 & 26.89 &
02 & typ\\
9 & 1 & 10.0 & 19.2 & 19.5 & 32.18 & 103.28 & 117.11 & 32.49 &
-0.1 & -0.6 & 0.4 & \nodata & \nodata &
44.06 & 44 & inc\\
11 & 1 & 3.4 & 6.5 & 8.7 & 31.84 & 696.29 & 47.67 & 696.29 & 2.7 &
0.0 & 6.2 & 55.03 & 74.06 & 124.41 &
01 & stp\\
23 & 1 & 2.4 & 5.5 & 14.9 & 31.98 & 60.34 & 72.48 & 54.54 & 1.2 &
8.4 & 0.6 & 4.15 & 21.22 & 45.93 & 13 &
typ\\
27 & 1 & 15.4 & 27.9 & 22.0 & 30.57 & 40.62 & 3.32 & 31.33 & 2.3 &
2.5 & 0.5 & 28.24 & 35.81 & 61.87 & 01
& inc\\
27 & 2 & 21.0 & 70.6 & 123.3 & 30.69 & 33.65 & 4.87 & 26.69 & 0.6
& -0.0 & 0.6 & 42.79 & 74.71 & 248.25 &
01 & stp\\
28 & 1 & 208.4 & 30.9 & 40.8 & 32.05 & 67.31 & 86.29 & 41.78 & 0.5
& 0.7 & -2.9 & 25.03 & 33.07 & 107.57
& 02 & inc\\
43 & 1 & 50.6 & 55.5 & 55.0 & 31.18 & 59.18 & 11.96 & 41.78 & 1.4
& 4.2 & 0.8 & 50.72 & 51.20 & 146.19 &
03 & srf\\
\enddata
\end{deluxetable}

\clearpage \clearpage

\begin{deluxetable}{ccccccccccc}
\centering \rotate \tabletypesize{\tiny} \tablewidth{0pt}
\tablecolumns{11} \tablecaption{Time-Resolved Spectroscopy
Analysis \label{tbl_TRS_results}}

\tablehead{

\colhead{Source} & \colhead{FlareNo} & \colhead{Type} &
\colhead{${\rm SegNum}$} & \colhead{${\rm <N>_{seg}}$} &
\colhead{${\rm N_{char}}$} & \colhead{T$_{char}$} &
\colhead{T$_{pk,TRS,noCH}$} & \colhead{T$_{pk,TRS,CH}$} &
\colhead{$\zeta_{TRS,noCH}$} & \colhead{$\zeta_{TRS,CH}$}\\

&&&&(cnts)&(cnts)&(MK)&(MK)&(MK)&&\\

(1)&(2)&(3)&(4)&(5)&(6)&(7)&(8)&(9)&(10)&(11)}

\startdata
   7 & 1 & typ &  8 &  202 & 1458 & 12.69 &  37.11 &  37.48 &  1.14 &  0.95\\
   9 & 1 & inc &  9 &  261 &  484 & 16.02 &  66.01 &  66.54 & -0.10 & -0.12\\
  11 & 1 & stp &  6 &  152 &  194 & 12.21 & 927.22 & 927.22 &  2.13 &  2.09\\
  23 & 1 & typ & 12 &  333 & 2090 & 16.16 &  53.72 &  54.10 &  1.15 &  0.99\\
  27 & 1 & inc &  3 &   80 &  175 & 12.10 &  32.50 &  33.61 &  1.51 &  1.38\\
  27 & 2 & stp & 11 &  130 &  175 & 12.10 &  45.83 &  50.22 &  0.67 & -0.49\\
  28 & 1 & inc & 22 &  586 &  381 & 15.76 &  59.45 &  59.48 &  0.69 &  0.67\\
  43 & 1 & srf &  8 &  208 &  210 & 15.96 &  74.90 &  77.30 &  2.09 &  2.15\\
\enddata

\tablecomments{Columns 1-3: Source number, flare number and flare
morphological type (reproduced from Table \ref{tbl_flare_prop2}).
Columns 4-5: Number of segments and the average number of X-ray
counts per segment used in the TRS analysis. Columns 6-7: Number
of X-ray counts in the ``characteristic'' segment and inferred
``characteristic'' plasma temperature. Columns 8-9: Inferred peak
flare plasma temperatures without (``noCH'') and with (``CH'')
account for the ``characteristic'' background. Columns 10-11:
Inferred average slopes of the temperature-density trajectory
without (``noCH'') and with (``CH'') accounting for the
``characteristic'' background.}

\end{deluxetable}

\clearpage \clearpage

\begin{deluxetable}{ccccccccccccccccc}
\centering \rotate \tabletypesize{\tiny} \tablewidth{0pt}
\tablecolumns{17} \tablecaption{COUP star properties
\label{tbl_source_prop}} \tablehead{

\colhead{Source} & \colhead{NC} & \colhead{BG} & \colhead{ME} &
\colhead{$\log N_H$} & \colhead{$J$} & \colhead{$H$} &
\colhead{$K_s$} & \colhead{Fl} & \colhead{$\Delta(H-K_s)$} &
\colhead{[3.6]} & \colhead{[4.5]} & \colhead{$M$} &
\colhead{$R$} & \colhead{$P$} & \colhead{$R_{cor}$} & \colhead{${\rm EW(Ca~{\tiny II})}$}\\

&(cnts)&(cnts)&(keV)&(cm$^{-2}$)&(mag)&(mag)&(mag)&&(mag)&(mag)&(mag)&($M_{\odot}$)&($10^{10}
$~cm)&(day)&($10^{10}$~cm)&(\AA)\\

(1)&(2)&(3)&(4)&(5)&(6)&(7)&(8)&(9)&(10)&(11)&(12)&(13)&(14)&(15)&(16)&(17)}

\startdata
       7 & 41952 & 362 & 1.20 & 20.93 &  8.85 &  8.10 &  7.95 & AAA000 & -0.28 & \nodata & \nodata & 2.12 &
43.61 & \nodata & \nodata & \nodata\\
       9 & 16271 & 199 & 1.34 & 21.37 & 10.22 &  9.65 &  9.46 & AAA000 & -0.15 & \nodata & \nodata & 2.11 &
20.72 & \nodata & \nodata & \nodata\\
      11 &  5700 & 123 & 1.53 & 21.69 & 10.53 &  9.46 &  8.60 & AAA000 &  0.25 & \nodata & \nodata & 0.69 &
24.22 & \nodata & \nodata & -14.60\\
      23 & 55449 & 320 & 1.29 & 21.24 & 10.01 &  9.33 &  9.09 & AAA000 & -0.16 & $ 8.873\pm 0.003$ & $
8.817\pm 0.002$ & 2.17 & 26.74 &  3.45 &  86.69 & \nodata\\
      27 &  5948 & 153 & 1.26 & 21.08 & 12.16 & 11.37 & 11.05 & AAA000 & -0.14 & $10.855\pm 0.005$ &
$10.789\pm 0.004$ & 0.53 & 12.67 &  1.50 &  31.10 &   1.80\\
      28 & 20863 & 149 & 1.39 & 21.10 & 11.53 & 10.84 & 10.53 & AAA000 & -0.09 & $10.332\pm 0.004$ &
$10.264\pm 0.004$ & 0.53 & 16.10 &  4.41 &  63.83 &   1.60\\
      43 &  7021 &  63 & 1.32 & 21.27 & 11.23 & 10.38 & 10.08 & AAA000 & -0.19 & $ 9.781\pm 0.005$ & $
9.716\pm 0.005$ & 0.40 & 20.44 & \nodata & \nodata &   1.40\\
      57 &  4093 &   5 & 1.13 & 20.68 & 11.01 & 10.53 & 10.28 & AEE000 & \nodata & $10.016\pm 0.003$ & $
9.988\pm 0.004$ & 1.19 & 15.19 &  2.34 &  54.78 &   1.60\\
\enddata

\tablecomments{Columns 1-5: COUP source number, source net counts,
background counts within source's extraction region, source X-ray
median energy, and inferred column density.  All values are from
\citet{Getman05}. Columns 6-8: VLT and 2MASS $JHK_s$ magnitudes as
reported in \citet{Getman05}. Column 9: Flag indicating source of
the NIR photometry (``\nodata''$=$VLT, otherwise 2MASS) and
quality of the 2MASS photometry as discussed in \citet{Getman05}.
Column 10: Color excess measured from the rightmost reddening
vector on the $J-H$ vs. $H-K_s$ color-color diagram (originating
at $\sim 0.1$~M$_{\odot}$ assuming an age of 1~Myr and using PMS
models of \citet{Siess00}). Columns 11-12: $Spitzer$-IRAC [3.6]
and [4.5] magnitudes and their statistical errors obtained in this
work. Columns 13-14: Stellar mass and radius as listed in
\citet{Getman05}. Column 15: Stellar rotation periods with
extended rotation data for COUP sources as listed in
\citet{Flaccomio05}. Column 16: Calculated Keplerian corotation
radius for stars with known $M$ and $P$. Column 17. Equivalent
width of the 8542~\AA\ Ca~II line as listed in \citet{Getman05}.}

\end{deluxetable}

\clearpage \clearpage

\clearpage

\clearpage

\clearpage

\clearpage

\clearpage

\clearpage

\clearpage

\clearpage

\clearpage

\clearpage

\clearpage

\clearpage

\clearpage

\clearpage

\clearpage

\end{document}